\documentclass[final,5p,times,twocolumn]{elsarticle}

\usepackage{amssymb}
\usepackage{amsmath}

\usepackage[ruled]{algorithm2e}
\usepackage{graphicx,color}
\usepackage{tikz}
\usepackage{pgfplots}
\usepackage{multirow}
\usepackage{hyperref}
\usepackage{acronym}
\usepackage{microtype}
\acrodef{avrmccc}[A-VRMCCC]{augmented Volterra recursive maximum complex correntropy criterion}
\acrodef{bc}[BC]{bias-compensated}
\acrodef{crlb}[CRLB]{Cram\'{e}r-Rao lower bound}
\acrodef{fim}[FIM]{Fisher information matrix}
\acrodef{fir}[FIR]{finite impulse response}
\acrodef{iq}[IQ]{in-phase and quadrature-phase}
\acrodef{ise}[ISE]{instantaneous squared error}
\acrodef{lut}[LUT]{lookup table}
\acrodef{lti}[LTI]{linear and time-invariant}
\acrodef{mmse}[MMSE]{minimum mean squared error}
\acrodef{mse}[MSE]{mean squared error}
\acrodef{miso}[MISO]{multiple-input-single-output}
\acrodef{nn}[NN]{neural network}
\acrodef{pa}[PA]{power amplifier}
\acrodef{pdf}[PDF]{probability density function}
\acrodef{rv}[RV]{real-valued}
\acrodef{siso}[SISO]{single-input-single-output}
\acrodef{wss}[WSS]{wide sense stationary}

\acrodef{kf}[KF]{Kalman filter}
\acrodef{aclkf1}[AC-LKF]{augmented complex-valued linear Kalman filter}
\acrodef{aclkf2}[AC-LKF]{augmented complex-valued linear Kalman filter}
\acrodef{aclkf}[AC-LKF]{augmented complex-valued linear \acs{kf}}

\acrodef{ekf1}[EKF]{extended Kalman filter}
\acrodef{ekf}[EKF]{extended \acs{kf}}
\acrodef{acekf1}[AC-EKF]{augmented complex-valued extended Kalman filter}
\acrodef{acekf2}[AC-EKF]{augmented complex-valued extended Kalman filter}
\acrodef{acekf}[AC-EKF]{augmented complex-valued extended \acs{kf}}

\acrodef{lms}[LMS]{least mean squares}
\acrodef{ls}[LS]{least squares}

\acrodef{nlms1}[NLMS]{normalized least mean squares}
\acrodef{nlms}[NLMS]{normalized \ac{lms}}

\acrodef{rls}[RLS]{recursive least squares}

\acrodef{wf}[WF]{Wiener filter}

\acrodef{2rblms1}[$2\mathbb{R}$-BLMS]{$2\mathbb{R}$ bilinear least mean squares}
\acrodef{2rblms2}[$2\mathbb{R}$-BLMS]{$2\mathbb{R}$ bilinear least mean squares}
\acrodef{2rblms}[$2\mathbb{R}$-BLMS]{$2\mathbb{R}$ bilinear \acs{lms}}

\acrodef{2rbnlms1}[$2\mathbb{R}$-BNLMS]{$2\mathbb{R}$ bilinear normalized least mean squares}
\acrodef{2rbnlms2}[$2\mathbb{R}$-BNLMS]{$2\mathbb{R}$ bilinear normalized least mean squares}
\acrodef{2rbnlms3}[$2\mathbb{R}$-BNLMS]{$2\mathbb{R}$ bilinear normalized \acs{lms}}
\acrodef{2rbnlms}[$2\mathbb{R}$-BNLMS]{$2\mathbb{R}$ bilinear \acs{nlms}}

\acrodef{4rblms1}[$4\mathbb{R}$-BLMS]{$4\mathbb{R}$ bilinear least mean squares}
\acrodef{4rblms2}[$4\mathbb{R}$-BLMS]{$4\mathbb{R}$ bilinear least mean squares}
\acrodef{4rblms}[$4\mathbb{R}$-BLMS]{$4\mathbb{R}$ bilinear \acs{lms}}

\acrodef{4rbnlms1}[$4\mathbb{R}$-BNLMS]{$4\mathbb{R}$ bilinear normalized least mean squares}
\acrodef{4rbnlms2}[$4\mathbb{R}$-BNLMS]{$4\mathbb{R}$ bilinear normalized least mean squares}
\acrodef{4rbnlms3}[$4\mathbb{R}$-BNLMS]{$4\mathbb{R}$ bilinear normalized \acs{lms}}
\acrodef{4rbnlms}[$4\mathbb{R}$-BNLMS]{$4\mathbb{R}$ bilinear \acs{nlms}}

\acrodef{bkf1}[BKF]{bilinear Kalman filter}
\acrodef{bkf2}[BKF]{bilinear Kalman filter}
\acrodef{bkf}[BKF]{bilinear \acs{kf}}
\acrodef{acbkf1}[AC-BKF]{augmented complex-valued bilinear Kalman filter}
\acrodef{acbkf2}[AC-BKF]{augmented complex-valued bilinear Kalman filter}
\acrodef{acbkf3}[AC-BKF]{augmented complex-valued bilinear Kalman filter}
\acrodef{acbkf4}[AC-BKF]{augmented complex-valued bilinear \acs{kf}}
\acrodef{acbkf}[AC-BKF]{augmented complex-valued \acs{bkf}}

\acrodef{cblms1}[C-BLMS]{complex-valued bilinear least mean squares}
\acrodef{cblms2}[C-BLMS]{complex-valued bilinear least mean squares}
\acrodef{cblms3}[C-BLMS]{complex-valued bilinear least mean squares}
\acrodef{cblms}[C-BLMS]{complex-valued bilinear \acs{lms}}

\acrodef{cbls1}[C-BLS]{complex-valued bilinear least squares}
\acrodef{cbls2}[C-BLS]{complex-valued bilinear least squares}
\acrodef{cbls3}[C-BLS]{complex-valued bilinear least squares}
\acrodef{cbls}[C-BLS]{complex-valued bilinear \acs{ls}}

\acrodef{cbnlms1}[C-BNLMS]{complex-valued bilinear normalized least mean squares}
\acrodef{cbnlms2}[C-BNLMS]{complex-valued bilinear normalized least mean squares}
\acrodef{cbnlms3}[C-BNLMS]{complex-valued bilinear normalized least mean squares}
\acrodef{cbnlms}[C-BNLMS]{complex-valued bilinear normalized \acs{lms}}

\acrodef{cbrls1}[C-BRLS]{complex-valued bilinear recursive least squares}
\acrodef{cbrls2}[C-BRLS]{complex-valued bilinear recursive least squares}
\acrodef{cbrls3}[C-BRLS]{complex-valued bilinear recursive least squares}
\acrodef{cbrls}[C-BRLS]{complex-valued bilinear \acs{rls}}

\acrodef{cbwf1}[C-BWF]{complex-valued bilinear Wiener filter}
\acrodef{cbwf2}[C-BWF]{complex-valued bilinear Wiener filter}
\acrodef{cbwf3}[C-BWF]{complex-valued bilinear Wiener filter}
\acrodef{cbwf}[C-BWF]{complex-valued bilinear \acs{wf}}

\acrodef{crblms1}[CR-BLMS]{mixed complex-valued-real-valued bilinear least mean squares}
\acrodef{crblms}[CR-BLMS]{mixed complex-valued-real-valued bilinear \acs{lms}}

\acrodef{crbls1}[CR-BLS]{mixed complex-valued-real-valued bilinear least squares}
\acrodef{crbls}[CR-BLS]{mixed complex-valued-real-valued bilinear \acs{ls}}

\acrodef{crbnlms1}[CR-BNLMS]{mixed complex-valued-real-valued bilinear normalized least mean squares}
\acrodef{crbnlms}[CR-BNLMS]{mixed complex-valued-real-valued bilinear \acs{nlms}}

\acrodef{crbrls1}[CR-BRLS]{mixed complex-valued-real-valued bilinear recursive least squares}
\acrodef{crbrls}[CR-BRLS]{mixed complex-valued-real-valued bilinear \acs{rls}}

\acrodef{crbwf1}[CR-BWF]{mixed complex-valued-real-valued bilinear Wiener filter}
\acrodef{crbwf}[CR-BWF]{mixed complex-valued-real-valued bilinear \acs{wf}}

\acrodef{hsaf1}[HSAF]{Hammerstein spline adaptive filter}
\acrodef{hsaf}[HSAF]{Hammerstein \ac{saf}}
\acrodef{saf}[SAF]{spline adaptive filter}
\acrodef{wsaf1}[WSAF]{Wiener spline adaptive filter}
\acrodef{wsaf}[WSAF]{Wiener \ac{saf}}

\acrodef{lse}[LSE]{least squares estimator}
\acrodef{cllse1}[C-LLSE]{complex-valued linear least squares estimator}
\acrodef{cllse2}[C-LLSE]{complex-valued linear least squares estimator}
\acrodef{cllse}[C-LLSE]{complex-valued linear \acs{lse}}

\acrodef{mle}[MLE]{maximum likelihood estimator}
\acrodef{clmle1}[C-LMLE]{complex-valued linear maximum likelihood estimator}
\acrodef{clmle2}[C-LMLE]{complex-valued linear maximum likelihood estimator}
\acrodef{clmle}[C-LMLE]{complex-valued linear \acs{mle}}

\acrodef{map}[MAP]{maximum a posteriori}

\acrodef{mmse}[MMSE]{minimum mean squared error}

\acrodef{cllms}[C-LLMS]{complex-valued linear least mean squares filter}

\acrodef{clnlms}[C-LNLMS]{complex-valued linear \acs{nlms} filter}

\acrodef{cblse1}[C-BLSE]{complex-valued bilinear least squares estimator}
\acrodef{cblse2}[C-BLSE]{complex-valued bilinear least squares estimator}
\acrodef{cblse3}[C-BLSE]{complex-valued bilinear least squares estimator}
\acrodef{cblse}[C-BLSE]{complex-valued bilinear \acs{lse}}

\acrodef{cbmle1}[C-BMLE]{complex-valued bilinear maximum likelihood estimator}
\acrodef{cbmle2}[C-BMLE]{complex-valued bilinear maximum likelihood estimator}
\acrodef{cbmle3}[C-BMLE]{complex-valued bilinear maximum likelihood estimator}
\acrodef{cbmle}[C-BMLE]{complex-valued bilinear \acs{mle}}

\newcommand{\ve}{\mathbf}
\newcommand{\m}{\mathbf}
\biboptions{sort&compress}

\journal{Signal Processing Open}

\begin{document}

\begin{frontmatter}



\title{Optimum and Adaptive Complex-Valued Bilinear Filters} 


\author[label1,label2]{Bernhard Plaimer}
\author[label1]{Matthias Wagner}
\author[label1]{Oliver Lang}
\author[label1]{and Mario Huemer} 

\affiliation[label1]{
	organization={Institute of Signal Processing, Johannes Kepler University Linz},
	addressline={Altenberger Straße 69},
    city={Linz},
    postcode={4040},
	state={Upper Austria},
    country={Austria}}
\affiliation[label2]{
	Corresponding author:,
	email={bernhard.plaimer@jku.at}}

\begin{abstract}
The identification of nonlinear systems is a frequent task in digital signal processing. Such nonlinear systems may be grouped into many sub-classes, whereby numerous nonlinear real-world systems can be approximated as bilinear models. Therefore, various optimum and adaptive bilinear filters have been introduced in recent years. Moreover, in many applications such as communications and radar, complex-valued bilinear systems in combination with complex-valued signals may occur.

Hence, in this work, we investigate the extension of real-valued bilinear filters to complex-valued bilinear filters. First, we derive complex-valued bilinear filters by applying two or four real-valued bilinear filters, and compare them with respect to their computational complexity and performance. Second, we introduce novel fully complex-valued bilinear filters, such as the \ac{cbwf3}, the \ac{cbls3} filter, the \ac{cblms3} filter, the \ac{cbnlms3} filter, and the \ac{cbrls3} filter. Finally, these filters are applied to identify complex-valued \ac{miso} systems and complex-valued Hammerstein models.
\end{abstract}



\begin{keyword}
Adaptive algorithm, bilinear filter, complex-valued, system identification.
\end{keyword}

\end{frontmatter}


\acresetall
\section{INTRODUCTION}
In many real-world applications, linear models provide sufficient descriptions of the input--output relations of unknown systems. Hence, many linear optimum and adaptive filters, such as the \ac{wf}, the \ac{ls} filter, the \ac{lms} filter, the \ac{nlms} filter, or the \ac{rls} filter are well established in the signal processing literature \cite{Kay_1993_1,Diniz_2008_1,Adali_2010_1}. However, many applications exhibit distinct nonlinear behavior, which leads to nonlinear models. For the identification of nonlinear systems, Volterra filters \cite{Diniz_2008_1,Volterra_1959_1}, \acp{nn} \cite{Haykin_1999_1}, \acp{saf} \cite{Scarpiniti_2013_1} and many more have been introduced.

Since a number of nonlinear systems can be approximated by bilinear systems, considerable research on bilinear systems has been conducted \cite{dosSantos_2009_1,Forssen_1993_1,Hu_1994_1,Kuo_2005_1,Ma_1994_1,Zhu_2000_1,Benesty_2017_1,Bai_2006_1,Ciochina_2017_1,Dogariu_2018_1,Paleologu_2017_1,EliseiIliescu_2017_1,EliseiIliescu_2018_1,EliseiIliescu_2018_2,EliseiIliescu_2018_3,Dogariu_2018_2,Paleologu_2018_1,Benesty_2021_1,Wagner_2025_1}. Note that the bilinear term can be defined in two different ways. First, in \cite{dosSantos_2009_1,Forssen_1993_1,Hu_1994_1,Kuo_2005_1,Ma_1994_1,Zhu_2000_1} it appears with respect to the multiplication of input and output signals of systems. Second, in \cite{Benesty_2017_1,Bai_2006_1,Ciochina_2017_1,Dogariu_2018_1,Paleologu_2017_1,EliseiIliescu_2017_1,EliseiIliescu_2018_1,EliseiIliescu_2018_2,EliseiIliescu_2018_3,Dogariu_2018_2,Paleologu_2018_1,Benesty_2021_1,Wagner_2025_1} the bilinear term is defined concerning the multiplication of systems' coefficients. In general, optimum and adaptive filters tailored to the first notion of bilinearity cannot be applied directly to systems embodying the second. Therefore, recent research has focused on the latter, which has led to bilinear optimum and adaptive filters such as the bilinear \ac{wf} \cite{Benesty_2017_1}, the bilinear \ac{ls} filter \cite{Bai_2006_1}, the bilinear \ac{lms} filter \cite{Ciochina_2017_1,Dogariu_2018_1}, the bilinear \ac{nlms} filter \cite{Paleologu_2017_1}, and the bilinear \ac{rls} filter \cite{EliseiIliescu_2017_1,EliseiIliescu_2018_2,EliseiIliescu_2018_1,EliseiIliescu_2018_3}. As summarized in \autoref{tab:1}, the current state-of-the-art covers solely real-valued bilinear filters. 

In many applications, e.g., in communications, also complex-valued bilinear systems in combination with complex-valued signals may occur \cite{Gesbert_1996_1,Qian_2019_1}. However, most existing literature on complex-valued nonlinear adaptive filters, such as complex-valued versions of \acp{saf} \cite{Campo_2018_1,Campo_2021_1,Campo_2021_2,Paireder_2021_1,Shi_2023_1}, complex-valued Volterra filters \cite{CrespoCadenas_2017_1, Qian_2024_1}, complex-valued \acp{nn} \cite{Lee_2022_1}, or complex-valued adaptive filters utilizing two-dimensional piecewise linear surfaces \cite{Liu_2025_1}, focuses on identifying systems where the nonlinearity is with respect to the input signal of the unknown system. In contrast, the bilinearity definition used in this paper, is with respect to the parameters of an unknown system and not necessarily with respect to the input signal. Therefore, the direct application of these filters to a general complex-valued bilinear system is not possible.

\begin{table*}[!t]
	\caption{Overview of optimum and adaptive bilinear filters}
	\centering
	\begin{tabular}{c| c| c| c| c| c}
		\hline
		filter type	&	real-valued bilinear filter 		&	\multicolumn{4}{c}{complex-valued bilinear filter structures}	\\
		&									&	$2\mathbb{R}$		&	$4\mathbb{R}$	&	\multicolumn{1}{c}{\shortstack{fully\\complex-valued}}	&	\multicolumn{1}{|c}{\shortstack{mixed complex-valued-\\real-valued}}	\\
		\hline
		\ac{wf}		&	\cite{Benesty_2017_1}									&	cf. \autoref{sec:2RLMS}	&	cf. \autoref{sec:4RLMS}	&	\textbf{\autoref{sec:FCVBF}}	&		\textbf{\autoref{sec:mixed}}	\\

		\ac{ls}		&	\cite{Bai_2006_1}									&	cf. \autoref{sec:2RLMS}	&	cf. \autoref{sec:4RLMS}	&	\textbf{\autoref{sec:CVBLS}}	&		\textbf{\autoref{sec:mixed}}	\\

		\ac{lms}	&	\cite{Ciochina_2017_1,Dogariu_2018_1}	&	\textbf{\autoref{sec:2RLMS}}									&	\textbf{\autoref{sec:4RLMS}}										&	\textbf{\autoref{sec:CVBLMS}}	&	\textbf{\autoref{sec:mixed}}	\\

		\ac{nlms}	&	\cite{Paleologu_2017_1}									&	cf. \autoref{sec:2RLMS}	&	cf. \autoref{sec:4RLMS}	&	\textbf{\autoref{sec:CVBNLMS}}	&	\textbf{\ref{sec:CVRVBLNLMS}}		\\

		\ac{rls}	&	\cite{EliseiIliescu_2017_1,EliseiIliescu_2018_2,EliseiIliescu_2018_1,EliseiIliescu_2018_3}	& 	cf. \autoref{sec:2RLMS}	&	cf. \autoref{sec:4RLMS}	&	\textbf{\autoref{sec:CVBRLS}}	&	\textbf{\autoref{sec:mixed}}	\\
		\hline
	\end{tabular}
	\label{tab:1}
\end{table*}

Hence, in this work, we extend the aforementioned real-valued bilinear filters to the complex domain. Note that two main approaches can be distinguished for deriving complex-valued filters. The first one is the split-complex-valued approach, where the input signal is split into its real and imaginary part. These two separate real-valued signals are then fed to either two or four real-valued filters to approximate the real and the imaginary part of a desired signal, as presented for linear filters in \cite{Mandic_2009_1}. In \autoref{tab:1} and the remainder of this work, these methods are referred to as $2\mathbb{R}$ and $4\mathbb{R}$ filter structures, respectively. Obviously, the $4\mathbb{R}$ filter structure may deal with signals, which show a correlation between the real and the imaginary parts as opposed to the $2\mathbb{R}$ filter structure.
The second approach to derive complex-valued filters is to use actual complex-valued filter coefficients, which will be referred to as fully complex-valued filter structure in the sequel. These fully complex-valued filters may show a number of advantages\cite{Adali_2011_1,Mandic_2009_1}:
\begin{itemize}
	\item As the $4\mathbb{R}$ filters they may also cope with a correlation between the real and the imaginary parts of the signals.
	\item They can be implemented with lower complexity.
	\item They typically allow for a more compact mathematical representation.
	\item In contrast to complex-valued linear filters, not every fully complex-valued bilinear system can be represented by a $4\mathbb{R}$ bilinear filter.
\end{itemize}

In this work, after a brief introduction and analysis of the $2\mathbb{R}$ and $4\mathbb{R}$ bilinear filter structures, we introduce several novel fully complex-valued bilinear filters, such as the \ac{cbwf}, the \ac{cbls} filter, the \ac{cblms} filter, the \ac{cbnlms} filter, and the \ac{cbrls} filter. For the sake of completeness, the update equations for mixed complex-valued-real-valued bilinear filters, where one of the two coefficient vectors is assumed to be real-valued, are also derived. A comprehensive overview of already existing and in this paper derived optimum and adaptive bilinear filters can be seen in \autoref{tab:1}. In addition to the derivations of these filters, convergence analyses are carried out for both the \ac{cbwf} and the \ac{cblms}-based filters. In this work, we additionally present representative applications of the proposed filters in the form of complex-valued \ac{miso} systems as in \cite{Benesty_2017_1,Bai_2006_1,Ciochina_2017_1,Paleologu_2017_1,EliseiIliescu_2017_1,Dogariu_2018_2,Dogariu_2018_1,EliseiIliescu_2018_2,EliseiIliescu_2018_1,EliseiIliescu_2018_3,Paleologu_2018_1,Benesty_2021_1} and complex-valued Hammerstein models \cite{Scarpiniti_2018_1, Liu_2023_1}. In conclusion, the following points briefly summarize the contributions of this paper:
\begin{itemize}
	\item Extension of the $2\mathbb{R}$ and $4\mathbb{R}$ filter structures to complex-valued bilinear filters.
	\item Derivation and analyses of several fully complex-valued bilinear filters, including convergence analyses for the \ac{cbwf} and the \ac{cblms}-based filters.
	\item Comparison of the different filter structures regarding their computational complexity.
	\item Presentation of mixed complex-valued-real-valued bilinear filters.
	\item Simulation results, supporting the applicability and advantages of fully complex-valued bilinear filters.
\end{itemize}
Note that this paper is based on the master's thesis \cite{Plaimer_2023_1} and extends it by several details, derivations, and proofs.

The rest of this paper is structured as follows: In \autoref{sec:CVBSM}, the complex-valued bilinear model is introduced. Derivations for $2\mathbb{R}$ bilinear filters, $4\mathbb{R}$ bilinear filters, fully complex-valued bilinear filters, and mixed complex-valued-real-valued bilinear filters are provided in \autoref{sec:CVBFRV}. Simulation results are presented in \autoref{sec:SIM}, and finally, \autoref{sec:CON} concludes this work. \\
\textit{Notation and Definitions:}

Lowercase and uppercase boldface letters denote vectors and matrices, respectively. Underlined vectors indicate the complex-valued augmentations of vectors, e.g., $\underline{\ve{x}} = \begin{bmatrix}
	\ve{x}^T \,\, \ve{x}^{H}
\end{bmatrix}^T$, where $\left(\cdot\right)^T$ indicates the transposition, and $\left(\cdot\right)^H$ indicates the complex conjugate transposition. Furthermore, $\left(\cdot\right)^*$ indicates the complex conjugate, and $\operatorname{tr}\left[\cdot\right]$ indicates the trace of a matrix. To represent an identity matrix, $\m{I}^{m \times m}$ is used, while the zero matrix is denoted by $\m{0}^{m \times n}$. The superscripts indicate the dimensions. The expectation operator is denoted by $\operatorname{E}\left[\cdot\right]$, and the real and imaginary parts of a variable are indicated by $\operatorname{Re}\left[\cdot\right]$ and $\operatorname{Im}\left[\cdot\right]$, respectively. Furthermore, the imaginary unit is represented by $\text{j}$. The vectorization operator is defined as
\begin{align}
\ve{a} = \operatorname{vec}\left[\m{A}\right] = \operatorname{vec}\left[\begin{bmatrix}
	\ve{a}_1	&	\cdots	&	\ve{a}_M
\end{bmatrix}\right] = \begin{bmatrix}
	\ve{a}_1 \\
	\vdots \\
	\ve{a}_M
\end{bmatrix} \in \mathbb{C}^{LM} \text{,} \label{equ:vecop}
\end{align}
where $\ve{a}_m \in \mathbb{C}^{L}$ denotes the $m$th column of $\m{A} \in \mathbb{C}^{L \times M}$. The operator $\operatorname{mat}_M\left[\cdot\right]$ is the inverse of the vectorization operator as described in \eqref{equ:vecop} and is defined as
\begin{align}
	\m{A} = \operatorname{mat}_M\left[\ve{a}\right] = \operatorname{mat}_M\begin{bmatrix}
		\ve{a}_1 \\
		\vdots \\
		\ve{a}_M
	\end{bmatrix} = \begin{bmatrix}
		\ve{a}_1	&	\cdots	&	\ve{a}_M
	\end{bmatrix} \text{,}
\end{align}
where the vector $\ve{a}$ is split into $M$ sub-vectors $\ve{a}_m$ of length $L$. The operator $\otimes$ is used for the Kronecker product.

\section{COMPLEX-VALUED BILINEAR MODEL} \label{sec:CVBSM}
\begin{figure}[!t]
	\centering
	\includegraphics[width=\columnwidth]{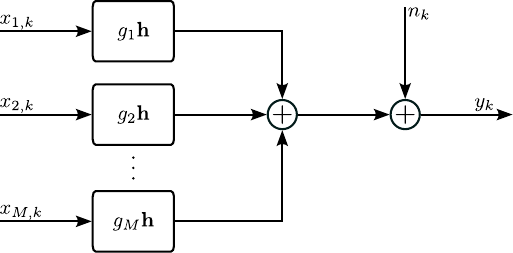}
	\caption{Block diagram of a complex-valued \ac{miso} system.}
	\label{fig:MISO}
\end{figure}

In this work, a complex-valued bilinear system is represented by the input--output relation
\begin{align}
	y_k = \ve{h}^H \m{X}_k \ve{g} + n_k \text{,} \label{equ:bi}
\end{align}
with $y_k \in \mathbb{C}$ as the output signal, $\m{X}_k \in \mathbb{C}^{L \times M}$ as the input signal matrix, and $n_k \in \mathbb{C}$ as complex-valued noise at a time instance $k$. The matrix $\m{X}_k$ depends on the values of the input signals, and its structure heavily depends on the application at hand. Likewise, the physical interpretation of the parameter vectors $\ve{h}$ and $\ve{g}$ is very application-dependent.
As an introductory toy example, the input--output relation of the complex-valued \ac{miso} system in \autoref{fig:MISO} can be written in the form of \eqref{equ:bi}. This model is valid when each channel impulse response is a scaled copy of a common impulse response $\ve{h}$. With $M$ input signals $x_{m,k}$ in combination with $M$ linear channels $g_m \ve{h} \in \mathbb{C}^L$, for $m = 1, \dots ,M$, the input signal matrix is defined as
\begin{align}
	\m{X}_k = \begin{bmatrix}
		\ve{x}_{1,k}	&	\cdots	&	\ve{x}_{M,k}
	\end{bmatrix} \text{,}
\end{align}
where the $m$th column $\ve{x}_{m,k}$ consists of the past $L$ input samples of the input signal $x_{m,k}$. The scalar $g_m \in \mathbb{C}$ is the $m$th entry of $\ve{g} \in \mathbb{C}^M$, and it may represent an amplitude and phase difference among the $M$ different channels.

\begin{figure}[!t]
	\centering
	\includegraphics[width=\columnwidth]{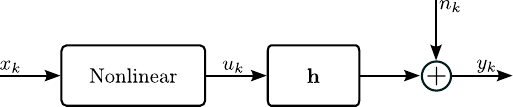}
	\caption{Block diagram of a complex-valued Hammerstein system.}
	\label{fig:Hammer}
\end{figure}

In a second example, consider a complex-valued Hammerstein system, as can be seen in \autoref{fig:Hammer}. Such systems arise, for example, in communication or radar transmitters affected by \ac{iq} imbalance and/or a nonlinear \ac{pa}. Assuming the nonlinearity is memory-free and linear in the parameters as
\begin{align}
	u_k = \sum_{m=1}^{M} g_m \varphi_m\left(x_k\right) \text{,}
\end{align}
with the input signal $x_k$, nonlinear functions $\varphi_m\left(x_k\right)$, and parameters $g_m$, the input--output relation of this complex-valued Hammerstein system again can be written in the form of \eqref{equ:bi}. The corresponding input signal matrix results in
\begin{align}
	\m{X}_k = \begin{bmatrix}
		\varphi_1\left(x_k\right)	&	\cdots	&	\varphi_M\left(x_k\right)  \\
		\vdots		&	\ddots	&	\vdots			\\
		\varphi_1\left(x_{k-L+1}\right)	&	\cdots	&	\varphi_M\left(x_{k-L+1}\right)
	\end{bmatrix} \text{.}
\end{align}
In both applications, the vectors $\m{h} \in \mathbb{C}^L$ and $\m{g} \in \mathbb{C}^M$ can be interpreted as coefficient vectors.

By analyzing \eqref{equ:bi}, it is easy to see that every bilinear model can be rewritten as a linear model as
\begin{align}
	y_k &= \ve{f}^T \widetilde{\ve{x}}_k +n_k \text{,} \label{equ:equiv}
\end{align}
where we denote $\ve{f} = \ve{g} \otimes \ve{h}^* \in \mathbb{C}^{L M}$ as the corresponding linear coefficient vector and $\widetilde{\ve{x}}_k = \operatorname{vec}\left[\m{X}_k\right] \in \mathbb{C}^{L M}$ as the corresponding input vector.

\begin{figure}[!t]
	\centering
	\includegraphics[width=\columnwidth]{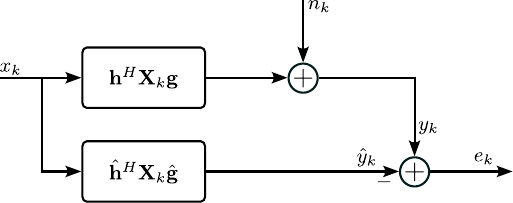}
	\caption{Block diagram of a system identification task using optimum or adaptive filters.}
	\label{fig:Adapt}
\end{figure}

Using optimum or adaptive filters, with their key structure illustrated in \autoref{fig:Adapt}, the goal is to find filter parameters $\hat{\ve{h}} \in \mathbb{C}^L$ and $\hat{\ve{g}} \in \mathbb{C}^M$ or $\hat{\ve{f}} \in \mathbb{C}^{LM}$ that minimize a cost function based on the error signal $e_k$.
While a bilinear system has $L+M$ unknown parameters, finding $\hat{\ve{f}}_k$ directly requires $LM$ parameters to be found.
Note that it is only possible to estimate $\ve{h}$ and $\ve{g}$ up to a complex-valued constant $\nu \in \mathbb{C}$, since
\begin{align}
	\ve{h}^H \m{X}_k \ve{g} = \left(\nu \ve{h}\right)^H \m{X}_k \ve{g} \frac{1}{\nu^*} \label{equ:amb} \text{,}
\end{align}
while $\ve{f}$ is unaffected by this scaling factor.

\section{COMPLEX-VALUED BILINEAR FILTERS}\label{sec:CVBFRV}
In the following chapter, several complex-valued bilinear filters are derived. To approximate the output of a system as in \eqref{equ:bi}, we investigate four different methods. First, this can be achieved by using either $2\mathbb{R}$ or $4\mathbb{R}$ bilinear filter structures. Second, a fully complex-valued bilinear filter structure is applied to estimate $\ve{h}$ and $\ve{g}$. The fourth method is based on the corresponding linear system shown in \eqref{equ:equiv}, thus a complex-valued linear filter may be used \cite{Adali_2011_1,Mandic_2009_1,Schreier_2010_1,Lang_2018_1}. However, this may result in slow convergence speeds because $L M$ coefficients have to be estimated instead of $L + M$ in the bilinear case.

It should be noted that $2\mathbb{R}$ bilinear filters and $4\mathbb{R}$ bilinear filters may be interpreted as simple extensions from the vast literature on real-valued bilinear filters \cite{Benesty_2017_1,Bai_2006_1,Ciochina_2017_1,Dogariu_2018_1,Paleologu_2017_1,EliseiIliescu_2017_1,EliseiIliescu_2018_2,EliseiIliescu_2018_1,EliseiIliescu_2018_3,Dogariu_2018_2,Paleologu_2018_1,Benesty_2021_1}. Thus, in this work only the derivations of the \ac{2rblms} filter and the \ac{4rblms} filter are shown. The derivations of further $2\mathbb{R}$ and $4\mathbb{R}$ optimum and adaptive bilinear filters can be carried out similarly.
\subsection{\ac{2rblms} filter} \label{sec:2RLMS}
\begin{figure}[!t]
	\centering
	\includegraphics[width=\columnwidth]{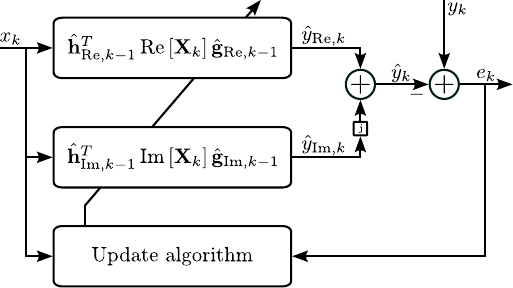}
	\caption{Block diagram of a $2\mathbb{R}$ bilinear filter structure.}
	\label{fig:SCV2R}
\end{figure}

The output of a \ac{2rblms} filter, as can be seen in \autoref{fig:SCV2R}, may be written as
\begin{align}
	\hat{y}_k &= \hat{\ve{h}}_{\operatorname{Re},k-1}^T \operatorname{Re}\left[\m{X}_k\right] \hat{\ve{g}}_{\operatorname{Re},k-1} + \label{equ:24} \\
	&\quad \,\, \text{j} \hat{\ve{h}}_{\operatorname{Im},k-1}^T \operatorname{Im}\left[\m{X}_k\right] \hat{\ve{g}}_{\operatorname{Im},k-1} \nonumber \\
	&=	\hat{y}_{\operatorname{Re},k} + \text{j} \hat{y}_{\operatorname{Im},k} \text{,}
\end{align}
with $\hat{\ve{h}}_{\operatorname{Re},k}, \hat{\ve{h}}_{\operatorname{Im},k} \in \mathbb{R}^L$, and $\hat{\ve{g}}_{\operatorname{Re},k}, \hat{\ve{g}}_{\operatorname{Im},k} \in \mathbb{R}^M$. Note that \eqref{equ:24} incorporates two real-valued bilinear \ac{lms} filters as discussed in \cite{Ciochina_2017_1} and intentionally excludes cross terms between the real and imaginary parts to provide the simplest extension of a real-valued bilinear filter. The \ac{4rblms} filter introduced later augments this structure by adding cross terms. Consequently, the \ac{2rblms} filter is a special case of the \ac{4rblms} filter with the cross terms set to zero.
With that, the error signal
\begin{align}
	e_{k} &= y_k - \hat{y}_{k} \\
	&= \operatorname{Re}\left[y_k\right] - \hat{y}_{\operatorname{Re},k} + \text{j}\left(\operatorname{Im}\left[y_k\right] - \hat{y}_{\operatorname{Im},k}\right)\\
	&= e_{\operatorname{Re},k} + \text{j} e_{\operatorname{Im},k} \label{equ:eRe}
\end{align}
and the \ac{ise} as the cost function
\begin{align}
	J_k &= e_k e_k^* \label{equ:costRe}
\end{align}
can be formulated.
Utilizing this cost function, the gradient descent based update equations for the filter coefficients follow as
\begin{align}
	\hat{\ve{h}}_{\operatorname{Re},k} &= \hat{\ve{h}}_{\operatorname{Re},k-1} - \mu_{\ve{h}_{\operatorname{Re}}} \frac{\partial J_k}{\partial \hat{\ve{h}}_{\operatorname{Re},k-1}} \text{,} \\
	\hat{\ve{h}}_{\operatorname{Im},k} &= \hat{\ve{h}}_{\operatorname{Im},k-1} - \mu_{\ve{h}_{\operatorname{Im}}} \frac{\partial J_k}{\partial \hat{\ve{h}}_{\operatorname{Im},k-1}}
\end{align}
and
\begin{align}
	\hat{\ve{g}}_{\operatorname{Re},k} &= \hat{\ve{g}}_{\operatorname{Re},k-1} - \mu_{\ve{g}_{\operatorname{Re}}} \frac{\partial J_k}{\partial \hat{\ve{g}}_{\operatorname{Re},k-1}} \text{,} \\
	\hat{\ve{g}}_{\operatorname{Im},k} &= \hat{\ve{g}}_{\operatorname{Im},k-1} - \mu_{\ve{g}_{\operatorname{Im}}} \frac{\partial J_k}{\partial \hat{\ve{g}}_{\operatorname{Im},k-1}} \text{,}
\end{align}
where $\mu_{\ve{h}_{\operatorname{Re}}}, \mu_{\ve{h}_{\operatorname{Im}}}, \mu_{\ve{g}_{\operatorname{Re}}}, \mu_{\ve{g}_{\operatorname{Im}}} \in \mathbb{R}$ are the step-sizes.
Inserting the gradients yields
\begin{align}
	\hat{\ve{h}}_{\operatorname{Re},k} &= \hat{\ve{h}}_{\operatorname{Re},k-1} + \mu_{\ve{h}_{\operatorname{Re}}} e_{\operatorname{Re},k} \operatorname{Re}\left[\m{X}_k\right] \hat{\ve{g}}_{\operatorname{Re},k-1}  \label{equ:012} \text{,} \\
	\hat{\ve{h}}_{\operatorname{Im},k} &= \hat{\ve{h}}_{\operatorname{Im},k-1} + \mu_{\ve{h}_{\operatorname{Im}}} e_{\operatorname{Im},k} \operatorname{Im}\left[\m{X}_k\right] \hat{\ve{g}}_{\operatorname{Im},k-1}  \label{equ:013} \text{.}
\end{align}
Furthermore, the update equations for $\hat{\ve{g}}_{\operatorname{Re},k}$ and $\hat{\ve{g}}_{\operatorname{Im},k}$ become
\begin{align}
	\hat{\ve{g}}_{\operatorname{Re},k} &= \hat{\ve{g}}_{\operatorname{Re},k-1} + \mu_{\ve{g}_{\operatorname{Re}}} e_{\operatorname{Re},k} \operatorname{Re}\left[\m{X}_k^T\right] \hat{\ve{h}}_{\operatorname{Re},k-1}  \label{equ:014} \text{,} \\
	\hat{\ve{g}}_{\operatorname{Im},k} &= \hat{\ve{g}}_{\operatorname{Im},k-1} + \mu_{\ve{g}_{\operatorname{Im}}} e_{\operatorname{Im},k} \operatorname{Im}\left[\m{X}_k^T\right] \hat{\ve{h}}_{\operatorname{Im},k-1}  \label{equ:015} \text{.}
\end{align}
\textit{Computational complexity:}

To compute \eqref{equ:012} and \eqref{equ:013}, $2 \left(L M + L + 1\right)$ real-valued scalar multiplications are necessary. Analogously, for \eqref{equ:014} and \eqref{equ:015}, $2 \left(L M + M + 1\right)$ real-valued scalar multiplications are needed.
When calculating the error $e_{\operatorname{Re},k}$, the first bilinear term in \eqref{equ:24} can be evaluated in two different ways by either starting with $\operatorname{Re}\left[\m{X}_k\right] \hat{\ve{g}}_{\operatorname{Re},k-1}$ or with $\hat{\ve{h}}_{\operatorname{Re},k-1}^T \operatorname{Re}\left[\m{X}_k\right]$. Similarly, also $e_{\operatorname{Im},k}$ can be calculated in two different ways. This yields $LM + L$ or $LM + M$ real-valued scalar multiplications, respectively. In total, this sums up to
\begin{align}
	N_{2 \mathbb{R},1} = 6 L M + 2 L + 4 M + 4 \label{equ:multi}
\end{align}
or
\begin{align}
	N_{2 \mathbb{R},2} = 6 L M + 4 L + 2 M + 4 \label{equ:multi1}
\end{align}
real-valued scalar multiplications. Additionally applying the Big O notation yields
\begin{align}
	N_{2 \mathbb{R}} = \mathcal{O}\left(L M\right) \text{.} \label{equ:multi2}
\end{align}
The results from \eqref{equ:multi}, \eqref{equ:multi1}, and \eqref{equ:multi2}  are summarized in \autoref{tab:complexity}.\\
\textit{Remark:}

\begin{table}[!t]
	\caption{Computational complexity of the complex-valued \acs{lms}-based filters}
	\centering
	\begin{tabular}{c|c|c}
		\hline
		filter type	&	\multicolumn{1}{c|}{\shortstack{\rule{0pt}{2ex}amount of real-valued\\scalar multiplications}}	&	Big O notation	\\
		\hline
		\multirow{2}{*}{\acs{2rblms}}			&	$6 L M + 2 L + 4 M + 4$	&	\multirow{2}{*}{$\mathcal{O}\left(L M\right)$}	\\	
		&	$6 L M + 4 L + 2 M + 4$	&	\\ \hline
		\multirow{2}{*}{\acs{4rblms}}			&	$12 L M + 4 L + 8 M + 8$	&	\multirow{2}{*}{$\mathcal{O}\left(L M\right)$}	\\	
		&	$12 L M + 8 L + 4 M + 8$	&	\\ \hline
		\multirow{2}{*}{\acs{cblms}}			&	$9 L M + 6 M + 3 L + 4$	&	\multirow{2}{*}{$\mathcal{O}\left(L M\right)$}	\\	
		&	$9 L M + 3 M + 6 L + 4$	&	\\ \hline
		\acs{cllms}	&	$6 LM + 2$	&	$\mathcal{O}\left(L M\right)$	\\
		\hline
	\end{tabular}
	\label{tab:complexity}
\end{table}

To identify a system that introduces a correlation between the real and the imaginary part of the output signal, it is well known from linear filters that this method may perform poorly \cite{Adali_2011_1,Mandic_2009_1}. The utilization of \ac{4rblms} filters may circumvent this problem as discussed in the following.

\subsection{\ac{4rblms} filter}\label{sec:4RLMS}
\begin{figure}[!t]
	\centering
	\includegraphics[width=\columnwidth]{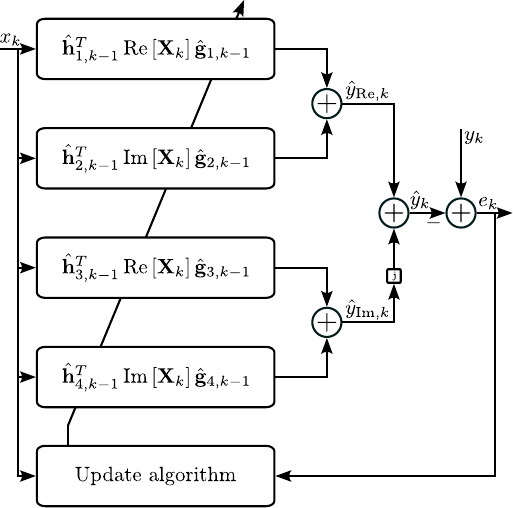}
	\caption{Block diagram of a $4\mathbb{R}$ bilinear filter structure.}
	\label{fig:SCV4R}
\end{figure}

When using four real-valued bilinear filters, as shown in \autoref{fig:SCV4R}, the filtered output becomes
\begin{align}
	\hat{y}_k &= \hat{\ve{h}}_{1,k-1}^T \operatorname{Re}\left[\m{X}_k\right] \hat{\ve{g}}_{1,k-1} + \\
	&\quad \, \, \hat{\ve{h}}_{2,k-1}^T \operatorname{Im}\left[\m{X}_k\right] \hat{\ve{g}}_{2,k-1} + \nonumber \\
	& \quad \, \, \text{j} \left(\hat{\ve{h}}_{3,k-1}^T \operatorname{Re}\left[\m{X}_k\right] \hat{\ve{g}}_{3,k-1} + \hat{\ve{h}}_{4,k-1}^T \operatorname{Im}\left[\m{X}_k\right] \hat{\ve{g}}_{4,k-1}\right) \nonumber \\
	&=	\hat{y}_{\operatorname{Re},k} + \text{j} \hat{y}_{\operatorname{Im},k} \text{,} \label{equ:est}
\end{align}
with $\hat{\ve{h}}_{l,k} \in \mathbb{R}^L$ and $\hat{\ve{g}}_{l,k} \in \mathbb{R}^M$ where $l = 1, \dots ,4$.
The same error signal and cost function as in \eqref{equ:eRe} and \eqref{equ:costRe} are used but with the estimated output \eqref{equ:est}.
Again using the gradient descent technique with the step-sizes $\mu_{\ve{h}_i} \in \mathbb{R}$ and $\mu_{\ve{g}_i} \in \mathbb{R}$ the update equations become
\begin{align}
	\hat{\ve{h}}_{i,k} &= \hat{\ve{h}}_{i,k-1} - \mu_{\ve{h}_i} \frac{\partial J_k}{\partial \hat{\ve{h}}_{i,k-1}} \text{,} \\
	\hat{\ve{g}}_{i,k} &= \hat{\ve{g}}_{i,k-1} - \mu_{\ve{g}_i} \frac{\partial J_k}{\partial \hat{\ve{g}}_{i,k-1}} \text{,}
\end{align}
for $i = 1, \dots ,4$.
Evaluating the gradients yields
\begin{align}
	\hat{\ve{h}}_{1,k} &= \hat{\ve{h}}_{1,k-1} + \mu_{\ve{h}_1} e_{\operatorname{Re},k} \operatorname{Re}\left[\m{X}_k\right] \hat{\ve{g}}_{1,k-1}  \label{equ:016} \text{,} \\
	\hat{\ve{h}}_{2,k} &= \hat{\ve{h}}_{2,k-1} + \mu_{\ve{h}_2} e_{\operatorname{Re},k} \operatorname{Im}\left[\m{X}_k\right] \hat{\ve{g}}_{2,k-1}  \label{equ:017} \text{,} \\
	\hat{\ve{h}}_{3,k} &= \hat{\ve{h}}_{3,k-1} + \mu_{\ve{h}_3} e_{\operatorname{Im},k} \operatorname{Re}\left[\m{X}_k\right] \hat{\ve{g}}_{3,k-1}  \label{equ:018} \text{,} \\
	\hat{\ve{h}}_{4,k} &= \hat{\ve{h}}_{4,k-1} + \mu_{\ve{h}_4} e_{\operatorname{Im},k} \operatorname{Im}\left[\m{X}_k\right] \hat{\ve{g}}_{4,k-1}  \label{equ:019} \text{,}
\end{align}
and
\begin{align}
	\hat{\ve{g}}_{1,k} &= \hat{\ve{g}}_{1,k-1} + \mu_{\ve{g}_1} e_{\operatorname{Re},k} \operatorname{Re}\left[\m{X}_k^T\right] \hat{\ve{h}}_{1,k-1}  \label{equ:020} \text{,} \\
	\hat{\ve{g}}_{2,k} &= \hat{\ve{g}}_{2,k-1} + \mu_{\ve{g}_2} e_{\operatorname{Re},k} \operatorname{Im}\left[\m{X}_k^T\right] \hat{\ve{h}}_{2,k-1}  \label{equ:021} \text{,} \\
	\hat{\ve{g}}_{3,k} &= \hat{\ve{g}}_{3,k-1} + \mu_{\ve{g}_3} e_{\operatorname{Im},k} \operatorname{Re}\left[\m{X}_k^T\right] \hat{\ve{h}}_{3,k-1}  \label{equ:022} \text{,} \\
	\hat{\ve{g}}_{4,k} &= \hat{\ve{g}}_{4,k-1} + \mu_{\ve{g}_4} e_{\operatorname{Im},k} \operatorname{Im}\left[\m{X}_k^T\right] \hat{\ve{h}}_{4,k-1}  \label{equ:023}  \text{.}
\end{align}
\textit{Computational complexity:}

For the calculations of the errors and the update equations
\begin{align}
	N_{4 \mathbb{R},1} = 2 N_{2 \mathbb{R},1} \label{equ:N4}
\end{align}
or
\begin{align}
	N_{4 \mathbb{R},2} = 2 N_{2 \mathbb{R},2} \label{equ:N41}
\end{align}
real-valued scalar multiplications are necessary. Similar to above, utilizing the Big O notation yields
\begin{align}
	N_{4 \mathbb{R}} = \mathcal{O}\left(L M\right) \text{.} \label{equ:N42}
\end{align}
The results from \eqref{equ:N4}, \eqref{equ:N41}, and \eqref{equ:N42} are summarized in \autoref{tab:complexity}. \\
\textit{Remark:}

Due to the use of four real-valued bilinear filters this method might be able to consider correlated signals, which, depending on the application, may yield a better performance. Still, there exist several reasons why the extension to the fully complex-valued case should be investigated. The fully complex-valued filters also alleviate the problem of the $2\mathbb{R}$ filter structure with correlated signals, they can be implemented with less computational complexity than the $4\mathbb{R}$ filter structure, and often the mathematical representation can be more compact \cite{Adali_2011_1,Mandic_2009_1}. Finally, it is worth emphasizing that many fully complex-valued bilinear systems cannot be modeled using four real-valued bilinear filters. Consider an arbitrary complex-valued bilinear system of the form \eqref{equ:bi}. Inserting $\ve{h} = \ve{h}_{\operatorname{Re}} + \text{j} \ve{h}_{\operatorname{Im}}$, $\ve{g} = \ve{g}_{\operatorname{Re}} + \text{j} \ve{g}_{\operatorname{Im}}$, and $\m{X}_k = \operatorname{Re}\left[\m{X}_k\right] + \text{j} \operatorname{Im}\left[\m{X}_k\right]$ into \eqref{equ:bi} yields
\begin{align}
	y_k = \,&\ve{h}_{\operatorname{Re}}^T \operatorname{Re}\left[\m{X}_k\right] \ve{g}_{\operatorname{Re}} - \ve{h}_{\operatorname{Re}}^T \operatorname{Im}\left[\m{X}_k\right] \ve{g}_{\operatorname{Im}} - \label{equ:8R}\\
	&\ve{h}_{\operatorname{Im}}^T \operatorname{Im}\left[\m{X}_k\right] \ve{g}_{\operatorname{Re}} - \ve{h}_{\operatorname{Im}}^T \operatorname{Re}\left[\m{X}_k\right] \ve{g}_{\operatorname{Im}} + \nonumber \\
	&\text{j} \left(\ve{h}_{\operatorname{Re}}^T \operatorname{Im}\left[\m{X}_k\right] \ve{g}_{\operatorname{Re}} + \ve{h}_{\operatorname{Re}}^T \operatorname{Re}\left[\m{X}_k\right] \ve{g}_{\operatorname{Im}} \right. + \nonumber \\
	&\ve{h}_{\operatorname{Im}}^T \operatorname{Re}\left[\m{X}_k\right] \ve{g}_{\operatorname{Re}} - \left.\ve{h}_{\operatorname{Im}}^T \operatorname{Im}\left[\m{X}_k\right] \ve{g}_{\operatorname{Im}}\right) + n_k \text{.} \nonumber
\end{align}
Since \eqref{equ:8R} consists of eight summands, it follows that a general fully complex-valued bilinear system cannot be modeled using $2\mathbb{R}$ and $4\mathbb{R}$ filter structures. Despite their limitations, $2\mathbb{R}$ and $4\mathbb{R}$ bilinear filters are frequently adequate in practice and therefore constitute a reasonable baseline for the subsequent filters. Utilizing a fully complex-valued filter structure resolves this issue.

\begin{figure}[!t]
	\centering
	\includegraphics[width=\columnwidth]{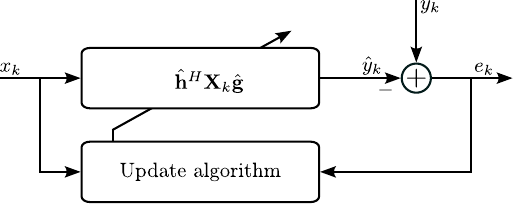}
	\caption{Block diagram of a fully complex-valued bilinear filter structure.}
	\label{fig:FCV}
\end{figure}

For all filters utilizing a fully complex-valued bilinear filter structure, as depicted in \autoref{fig:FCV}, it will be necessary to calculate derivatives of real-valued cost functions with respect to complex-valued coefficient vectors. This cannot be done in the classical sense, because real-valued functions are in general not holomorphic functions. This follows from the fact that they in general do not fulfill the Cauchy-Riemann differential equations \cite{Adali_2011_1,Mandic_2009_1}. Hence, we use Wirtinger's calculus \cite{Wirtinger_1927_1} to derive those gradients. These derivations can be done similarly as in \cite{Adali_2011_1,Schreier_2010_1,Mandic_2009_1,Lang_2018_1}.

\subsection{\ac{cbwf}} \label{sec:FCVBF}
Our investigations of fully complex-valued bilinear filters begin with the \ac{cbwf}.
Note that in this section only \ac{wss} signals are considered.

\acp{wf} use the \ac{mse} cost function
\begin{align}
	J &= \operatorname{E}\left[e_k e_k^*\right] \in \mathbb{R} \label{equ:002} \text{,}
\end{align}
with the error for the fully complex-valued bilinear filter structure given as
\begin{align}
	e_k = y_k - \hat{\ve{h}}^H \m{X}_k \hat{\ve{g}} \text{.}
\end{align}
Applying Wirtinger's calculus to derive the gradient of \eqref{equ:002} with respect to $\hat{\ve{h}}$ produces
\begin{align}
	\frac{\partial{J}}{\partial{\hat{\ve{h}}}} = - \m{R}_{\ve{X}y}^* \hat{\ve{g}}^* + \m{R}_{\ve{g}}^* \hat{\ve{h}}^* \text{,} \label{equ:003}
\end{align}
where $\m{R}_{\m{X}y} = \operatorname{E}\left[\m{X}_k y_k^*\right] \in \mathbb{C}^{L \times M}$ and $\m{R}_{\ve{g}} = \operatorname{E}\left[\m{X}_k \hat{\ve{g}} \hat{\ve{g}}^H \m{X}_k^H\right] \in \mathbb{C}^{L \times L}$. Setting \eqref{equ:003} to zero yields
\begin{align}
	\hat{\ve{h}} = \m{R}_{\ve{g}}^{-1} \m{R}_{\ve{X}y} \hat{\ve{g}} \label{equ:1} \text{.}
\end{align}
Similarly, the gradient of \eqref{equ:002} with respect to $\hat{\ve{g}}$ can be derived as
\begin{align}
	\frac{\partial{J}}{\partial{\hat{\ve{g}}}} = - \m{R}_{\ve{X}y}^T \hat{\ve{h}}^* + \m{R}_{\ve{h}}^* \hat{\ve{g}}^* \label{equ:004} \text{,}
\end{align}
with $\m{R}_{\ve{h}} = \operatorname{E}\left[\m{X}_k^H \hat{\ve{h}} \hat{\ve{h}}^H \m{X}_k\right] \in \mathbb{C}^{M \times M}$.
Setting \eqref{equ:004} to zero we obtain
\begin{align}
	\hat{\ve{g}} = \m{R}_{\ve{h}}^{-1} \m{R}_{\ve{X}y}^H \hat{\ve{h}} \label{equ:01} \text{.}
\end{align}
It is easy to see that \eqref{equ:1} and \eqref{equ:01} depend on each other. As in the real-valued case \cite{Benesty_2017_1}, this issue can be resolved by evaluating \eqref{equ:1} and \eqref{equ:01} in an alternating manner as
\begin{align}
	\hat{\ve{h}}_n = \m{R}_{\ve{g}, n-1}^{-1} \m{R}_{\ve{X}y} \hat{\ve{g}}_{n-1} \label{equ:001}
\end{align}
and
\begin{align}
	\hat{\ve{g}}_n = \m{R}_{\ve{h}, n}^{-1} \m{R}_{\ve{X}y}^H \hat{\ve{h}}_n \text{,} \label{equ:0001}
\end{align}
with $\hat{\ve{h}}_n \in \mathbb{C}^L$, $\hat{\ve{g}}_n \in \mathbb{C}^M$, $\m{R}_{\ve{g}, n-1} = \operatorname{E}\left[\m{X}_n \hat{\ve{g}}_{n-1} \hat{\ve{g}}_{n-1}^H \m{X}_n^H\right] \in \mathbb{C}^{L \times L}$, $\m{R}_{\ve{h}, n} = \operatorname{E}\left[\m{X}_n^H \hat{\ve{h}}_n \hat{\ve{h}}_n^H \m{X}_n\right] \in \mathbb{C}^{M \times M}$, and $n$ representing the iteration index. \\
\textit{Convergence:}

By investigating the cost function \eqref{equ:002}, it is possible to show that alternately evaluating \eqref{equ:001} and \eqref{equ:0001} cannot increase the \ac{mse}, leading to convergence. The proof for this can be found in \ref{app:wf}.\\
\textit{Remark:}

As shown similarly in \cite{Benesty_2017_1}, the matrices $\m{R}_{\ve{g},n-1}$ and $\m{R}_{\ve{h}, n}$ can be rewritten to
\begin{align}
	\m{R}_{\ve{g}, n-1} = (\hat{\ve{g}}_{n-1} \otimes \m{I}^{L \times L})^T \m{R}_{\widetilde{\ve{x}}\widetilde{\ve{x}}} (\hat{\ve{g}}_{n-1} \otimes \m{I}^{L \times L})^* \label{equ:10}
\end{align}
and
\begin{align}
	\m{R}_{\ve{h}, n} = (\m{I}^{M \times M} \otimes \hat{\ve{h}}_n)^T \m{R}_{\widetilde{\ve{x}}\widetilde{\ve{x}}}^*
	(\m{I}^{M \times M} \otimes \hat{\ve{h}}_n)^* \label{equ:11}
\end{align}
with the covariance matrix $\m{R}_{\widetilde{\ve{x}}\widetilde{\ve{x}}} = \operatorname{E}\left[\widetilde{\ve{x}}_k \widetilde{\ve{x}}_k^H\right] \in \mathbb{C}^{LM \times LM}$.
Hence, to evaluate the \ac{cbwf} it is necessary to know $\m{R}_{\widetilde{\ve{x}}\widetilde{\ve{x}}}$ and $\m{R}_{\ve{X}y}$. If these statistical quantities are not available, they may be estimated in advance by
\begin{align}
	\hat{\m{R}}_{\widetilde{\ve{x}}\widetilde{\ve{x}}} = \frac{1}{N} \sum_{i=1}^{N} \widetilde{\ve{x}}_i \widetilde{\ve{x}}_i^H \label{equ:12}
\end{align}
and
\begin{align}
	\hat{\ve{r}}_{\widetilde{\ve{x}}y} = \frac{1}{N} \sum_{i=1}^{N} \widetilde{\ve{x}}_i y_i^* \text{,}  \label{equ:13}
\end{align}
where $N$ is the number of samples used for the estimation and $\hat{\m{R}}_{\m{X}y} = \operatorname{mat}_M\left[\hat{\ve{r}}_{\widetilde{\ve{x}}y}\right]$.

Note that $\hat{\ve{g}}_{0} = \ve{0}$ is a bad choice for the initialization, because $\ve{h}_1$ would go towards infinity. If no prior knowledge about $\ve{g}$ or $\ve{h}$ is available, $\hat{\ve{g}}_{0}$ can be chosen randomly, but unequal to the zero vector. A final summary of the \ac{cbwf} is given in \autoref{alg:wfalg}.

A second way to handle unknown statistics is by implicitly estimating them using the \ac{ls} approach as shown in the following chapter.

\begin{algorithm}[t]
	\DontPrintSemicolon
	\emph{Initialize variables:}\;
	$\hat{\ve{g}}_{0} \ne \mathbf{0}$\;
	\For{$n = 1,2,3,\ldots$}{
		\emph{Parameter update of $\hat{\ve{h}}_{n}$:}\;
		$\m{R}_{\ve{g}, n-1} = (\hat{\ve{g}}_{n-1} \otimes \m{I}^{L \times L})^T \m{R}_{\widetilde{\ve{x}}\widetilde{\ve{x}}} (\hat{\ve{g}}_{n-1} \otimes \m{I}^{L \times L})^*$\;
		$\hat{\ve{h}}_n = \m{R}_{\ve{g}, n-1}^{-1} \m{R}_{\ve{X}y} \hat{\ve{g}}_{n-1}$\;
		\BlankLine		  \emph{Parameter update of $\hat{\ve{g}}_{n}$:}\;
		$\m{R}_{\ve{h}, n} = (\m{I}^{M \times M} \otimes \hat{\ve{h}}_n)^T \m{R}_{\widetilde{\ve{x}}\widetilde{\ve{x}}}^*
		(\m{I}^{M \times M} \otimes \hat{\ve{h}}_n)^*$\;
		$\hat{\ve{g}}_n = \m{R}_{\ve{h}, n}^{-1} \m{R}_{\ve{X}y}^H \hat{\ve{h}}_n$\;
	}
	\caption{\ac{cbwf}}
	\label{alg:wfalg}
\end{algorithm}

\subsection{\ac{cbls} filter}\label{sec:CVBLS}
Similarly to the complex-valued linear \ac{ls} filter, the \ac{cbls} filter does not require knowledge of any statistical signal properties. The corresponding cost function is
\begin{align}
	J = \sum_{i = 1}^{N} e_i e_i^* \label{equ:lse} \text{,}
\end{align}
where $N$ data-points $\{y_i, \m{X}_i\}$ have to be collected during a training phase. Similarly as for the \ac{cbwf}, the gradient of \eqref{equ:lse} with respect to $\hat{\ve{h}}$ produces
\begin{align}
	\frac{\partial{J}}{\partial{\hat{\ve{h}}}} = \sum_{i=1}^{N} -y_i \m{X}_i^* \hat{\ve{g}}^* + \m{X}_i^* \hat{\ve{g}}^* \hat{\ve{g}}^T \m{X}_i^T \hat{\ve{h}}^* \text{.}
\end{align}
Equating this gradient to zero yields
\begin{align}
	\hat{\ve{h}} = \left(\sum_{i=1}^{N} \m{X}_i \hat{\ve{g}} \hat{\ve{g}}^H \m{X}_i^H\right)^{-1} \sum_{i=1}^{N} y_i^* \m{X}_i \hat{\ve{g}} \text{.} \label{equ:lsh}
\end{align}
Analogously, the gradient of \eqref{equ:lse} with respect to $\hat{\ve{g}}$ is
\begin{align}
	\frac{\partial{J}}{\partial{\hat{\ve{g}}}} = \sum_{i=1}^{N} -y_i^* \m{X}_i^T \hat{\ve{h}}^* + \m{X}_i^T \hat{\ve{h}}^* \hat{\ve{h}}^T \m{X}_i^* \hat{\ve{g}}^* \text{,}
\end{align}
which yields
\begin{align}
	\hat{\ve{g}} = \left(\sum_{i=1}^{N} \m{X}_i^H \hat{\ve{h}} \hat{\ve{h}}^H \m{X}_i \right)^{-1} \sum_{i=1}^{N} y_i \m{X}_i^H \hat{\ve{h}} \text{.} \label{equ:lsg}
\end{align}
Again, alternately evaluating \eqref{equ:lsh} and \eqref{equ:lsg} leads to the final update equations
\begin{align}
	\hat{\ve{h}}_n &= \left(\sum_{i=1}^{N} \m{X}_i \hat{\ve{g}}_{n-1} \hat{\ve{g}}_{n-1}^H \m{X}_i^H\right)^{-1} \sum_{i=1}^{N} y_i^* \m{X}_i \hat{\ve{g}}_{n-1} \label{equ:25}
\end{align}
and
\begin{align}
	\hat{\ve{g}}_n &= \left(\sum_{i=1}^{N} \m{X}_i^H \hat{\ve{h}}_n \hat{\ve{h}}_n^H \m{X}_i \right)^{-1} \sum_{i=1}^{N} y_i \m{X}_i^H \hat{\ve{h}}_n \text{,} \label{equ:26}
\end{align}
depicted in \autoref{alg:lsalg}.\\
\textit{Remark:}

Similarly as for the \ac{cbwf}, by investigating the cost function \eqref{equ:lse}, it is possible to show that alternately evaluating \eqref{equ:25} and \eqref{equ:26} cannot increase the costs in each iteration, leading to convergence.

Furthermore, it is easy to show, that the \ac{cbls} filter is mathematically identical to the \ac{cbwf} when using the estimates \eqref{equ:12} and \eqref{equ:13} for $\m{R}_{\widetilde{\ve{x}}\widetilde{\ve{x}}}$ and $\ve{r}_{\widetilde{\ve{x}}y}$, respectively. 

When the underlying system is time-varying, an adaptive filter is advantageous to track parameter changes. Hence, a complex-valued bilinear \ac{rls} filter and a complex-valued bilinear \ac{lms} filter are derived in the following sections.

\begin{algorithm}[t]
	\DontPrintSemicolon
	\emph{Initialize variables:}\;
	$\hat{\ve{g}}_{0} \ne \mathbf{0}$\;
	\For{$n = 1,2,3,\ldots$}{
		\emph{Parameter update of $\hat{\ve{h}}_{n}$:}\;
		$\hat{\ve{h}}_n = \left(\sum_{i=1}^{N} \m{X}_i \hat{\ve{g}}_{n-1} \hat{\ve{g}}_{n-1}^H \m{X}_i^H\right)^{-1} \sum_{i=1}^{N} y_i^* \m{X}_i \hat{\ve{g}}_{n-1}$\;
		\BlankLine		  \emph{Parameter update of $\hat{\ve{g}}_{n}$:}\;
		$\hat{\ve{g}}_n = \left(\sum_{i=1}^{N} \m{X}_i^H \hat{\ve{h}}_n \hat{\ve{h}}_n^H \m{X}_i \right)^{-1} \sum_{i=1}^{N} y_i \m{X}_i^H \hat{\ve{h}}_n$\;
	}
	\caption{\ac{cbls} filter}
	\label{alg:lsalg}
\end{algorithm}

\subsection{\ac{cblms} filter}\label{sec:CVBLMS}
In this section, the \ac{cblms} filter is derived.
With the error signal
\begin{align}
	e_k &= y_k - \hat{y}_k \\
	&= y_k - \hat{\ve{h}}_{k-1}^H \m{X}_k \hat{\ve{g}}_{k-1} \label{equ:ebi} \text{,}
\end{align}
the \ac{ise}
\begin{align}
	J_k &= e_k e_k^* \label{equ:005}
\end{align}
is now applied as a cost function.
Application of a gradient descent method yields
\begin{align}
	\hat{\ve{h}}_k &= \hat{\ve{h}}_{k-1} - \mu_{\ve{h}} \frac{\partial J_k}{\partial \hat{\ve{h}}_{k-1}^*} \label{equ:uphl}
\end{align}
and
\begin{align}
	\hat{\ve{g}}_k &= \hat{\ve{g}}_{k-1} - \mu_{\ve{g}} \frac{\partial J_k}{\partial \hat{\ve{g}}_{k-1}^*} \text{.} \label{equ:upgl}
\end{align}
where $\mu_{\ve{h}}, \mu_{\ve{g}} \in \mathbb{R}$ are the step-sizes. Similarly as in \autoref{sec:FCVBF}, the gradients of \eqref{equ:005} with respect to $\hat{\ve{h}}_{k-1}^*$ and $\hat{\ve{g}}_{k-1}^*$ can be derived as
\begin{align}
	\frac{\partial J_k}{\partial \hat{\ve{h}}_{k-1}^*}	&= -e_k^* \m{X}_k \hat{\ve{g}}_{k-1} \label{equ:gradhl}
\end{align}
and
\begin{align}
	\frac{\partial J_k}{\partial \hat{\ve{g}}_{k-1}^*} &= -e_k \m{X}_k^H \hat{\ve{h}}_{k-1} \text{,} \label{equ:gradgl}
\end{align}
respectively.
Inserting \eqref{equ:gradhl} and \eqref{equ:gradgl} into \eqref{equ:uphl} and \eqref{equ:upgl}, delivers the final update equations
\begin{align}
	\hat{\ve{h}}_k &= \hat{\ve{h}}_{k-1} + \mu_{\ve{h}} e_k^* \m{X}_k \hat{\ve{g}}_{k-1} \label{equ:nuph}
\end{align}
and
\begin{align}
	\hat{\ve{g}}_k &= \hat{\ve{g}}_{k-1} + \mu_{\ve{g}} e_k \m{X}_k^H \hat{\ve{h}}_{k-1} \label{equ:nupg} \text{.}
\end{align}
\textit{Computational complexity:}

Exploiting the fact that multiplications of complex-valued scalars can be done with three real-valued scalar multiplications \cite{Malathi_2019_1}, it is possible to compute \eqref{equ:nuph} and \eqref{equ:nupg} with
\begin{align}
	N_{\mathbb{C},1} = 9 L M + 6 M + 3 L + 4 \label{equ:00001}
\end{align}
or
\begin{align}
	N_{\mathbb{C},2} = 9 L M + 3 M + 6 L + 4 \label{equ:00002}
\end{align}
real-valued scalar multiplications, which are less than in \eqref{equ:N4} and \eqref{equ:N41}. The Big O notation yields the same order of complexity \begin{align}
	N_{\mathbb{C}} = \mathcal{O}\left(L M\right) \text{.} \label{equ:00003}
\end{align}
However, for practical implementations on hardware, the reduction in the number of real-valued scalar multiplications is a significant advantage of the proposed \ac{cblms} filter.

In \autoref{tab:complexity}, we summarize the results from \eqref{equ:00001}, \eqref{equ:00002}, \eqref{equ:00003}, along with the computational complexity of the \ac{cllms} filter. The table shows that the \ac{cllms} filter requires fewer real-valued scalar multiplications than the \ac{cblms} filter, while both have the same Big-O order. However, this apparent advantage is tempered by the fact that the \ac{cllms} filter typically needs many more iterations to reach the same steady-state error as the \ac{cblms} (see \autoref{sec:SIM}). Moreover, the complexity advantage of the linear approach disappears for \ac{ls}-based filters or \ac{wf}-based algorithms, where matrix inversions dominate the computational complexity. The linear approaches scale as $\mathcal{O}\left(L^3 M^3\right)$ versus $\mathcal{O}\left(L^3 + M^3\right)$ for the bilinear approaches. \\
\textit{Convergence:}

As is shown in \ref{app:lms}, to ensure stability the step-sizes should follow the boundaries
\begin{align}
	0 < \mu_{\ve{h}} < \frac{2}{L \sigma_x^2 \operatorname{E}\left[||\hat{\ve{g}}_{k-1}||_2^2\right]}
\end{align}
and
\begin{align}
	0 < \mu_{\ve{g}} < \frac{2}{M \sigma_x^2 \operatorname{E}\left[||\hat{\ve{h}}_{k-1}||_2^2\right]} \text{.}
\end{align}
Furthermore, to guarantee convergence on the mean the step-sizes should be chosen such that
\begin{align}
	\Delta = 2 - \sigma_x^2\left(\mu_{\ve{h}} L \frac{1}{|\nu|^2} ||\ve{g}||_2^2 + \mu_{\ve{g}} M |\nu|^2 ||\ve{h}||_2^2\right) \label{equ:Delta}
\end{align}
is unequal to zero.

In \autoref{alg:lmsalg} the final \ac{cblms} filter is depicted. To improve convergence, the \ac{lms} algorithm is often extended with normalized step sizes. The next section presents this approach.

\begin{algorithm}[t]
	\DontPrintSemicolon
	\emph{Initialize variables:}\;
	$0 < \mu_{\ve{h}} < \frac{2}{L \sigma_x^2 \operatorname{E}\left[||\hat{\ve{g}}_{k-1}||_2^2\right]}$\;
	$0 < \mu_{\ve{g}} < \frac{2}{M \sigma_x^2 \operatorname{E}\left[||\hat{\ve{h}}_{k-1}||_2^2\right]}$\;
	With $\mu_{\ve{h}}$ and $\mu_{\ve{g}}$ such that $\Delta \ne 0$\;
	$\hat{\ve{h}}_{0} \ne \mathbf{0}$ and $\hat{\ve{g}}_{0} \ne \mathbf{0}$\;
	\For{$k = 1,2,3,\ldots$}{
		$e_k = y_k - \hat{\ve{h}}_{k-1}^H \m{X}_k \hat{\ve{g}}_{k-1}$\;
		\emph{Parameter update of $\hat{\ve{h}}_{k}$:}\;
		$\hat{\ve{h}}_k = \hat{\ve{h}}_{k-1} + \mu_{\ve{h}} e_k^* \m{X}_k \hat{\ve{g}}_{k-1}$\;
		\BlankLine		  \emph{Parameter update of $\hat{\ve{g}}_{k}$:}\;
		$\hat{\ve{g}}_k = \hat{\ve{g}}_{k-1} + \mu_{\ve{g}} e_k \m{X}_k^H \hat{\ve{h}}_{k-1}$\;
	}
	\caption{\ac{cblms} filter}
	\label{alg:lmsalg}
\end{algorithm}

\subsection{\ac{cbnlms} filter} \label{sec:CVBNLMS}
One way to derive the \ac{nlms} algorithm is to set the a posteriori error of the adaptive filter to zero, and to derive the step-size from this condition \cite{Paleologu_2017_1}. Having this in mind we introduce the a posteriori errors
\begin{align}
	\bar{e}_k &= y_k - \hat{\ve{h}}^H_k \m{X}_k \hat{\ve{g}}_{k-1} \label{equ:peg}
\end{align}
and
\begin{align}
	\tilde{e}_k &= y_k - \hat{\ve{h}}^H_{k-1} \m{X}_k \hat{\ve{g}}_{k} \label{equ:peh} \text{.}
\end{align}
Inserting \eqref{equ:nuph} and \eqref{equ:nupg} into \eqref{equ:peg} and \eqref{equ:peh} yields
\begin{align}
	\bar{e}_k &= e_k \left(1 - \mu_{\ve{h}} \hat{\ve{g}}_{k-1}^H \m{X}_k^H \m{X}_k \hat{\ve{g}}_{k-1}\right)
\end{align}
and
\begin{align}
	\tilde{e}_k &= e_k \left(1 - \mu_{\ve{g}} \hat{\ve{h}}^H_{k-1} \m{X}_k \m{X}_k^H \hat{\ve{h}}_{k-1}\right) \text{.}
\end{align}
Setting those a posteriori errors to zero, and assuming that $e_k \ne 0$ yields the step-sizes
\begin{align}
	\mu_{\ve{h},k} &= \frac{1}{\hat{\ve{g}}_{k-1}^H \m{X}_k^H \m{X}_k \hat{\ve{g}}_{k-1}} \label{equ:111}
\end{align}
and
\begin{align}
	\mu_{\ve{g},k} &= \frac{1}{\hat{\ve{h}}^H_{k-1} \m{X}_k \m{X}_k^H \hat{\ve{h}}_{k-1}} \text{.} \label{equ:112}
\end{align}
Finally, in accordance with the corresponding procedure for the ordinary \ac{nlms} algorithm, the update equations for the \ac{cbnlms} filter can be formulated as
\begin{align}
	\hat{\ve{h}}_k &= \hat{\ve{h}}_{k-1} + \frac{\alpha_{\ve{h}} \m{X}_k \hat{\ve{g}}_{k-1}}{\delta_{\ve{h}} + \hat{\ve{g}}_{k-1}^H \m{X}_k^H \m{X}_k \hat{\ve{g}}_{k-1}} e^*_k \label{equ:updh1}
\end{align}
and
\begin{align}
	\hat{\ve{g}}_k &= \hat{\ve{g}}_{k-1} + \frac{\alpha_{\ve{g}} \m{X}_k^H \hat{\ve{h}}_{k-1}}{\delta_{\ve{g}} + \hat{\ve{h}}^H_{k-1} \m{X}_k \m{X}_k^H \hat{\ve{h}}_{k-1}} e_k \text{,} \label{equ:updg1}
\end{align}
where $\alpha_{\ve{h}} \in \mathbb{R}$ and $\alpha_{\ve{g}} \in \mathbb{R}$ denote the normalized step-sizes and $\delta_{\ve{h}} \in \mathbb{R}$ and $\delta_{\ve{g}} \in \mathbb{R}$ are regularization constants to prevent divisions by very small numbers. As with the linear \ac{nlms} filter, the regularization constants should be small enough not to significantly affect the step sizes, yet large enough to prevent numerical issues. \\
\textit{Convergence:}

Without restricting to real-valued signals and using the instantaneous approximations $\hat{\ve{g}}_{k-1}^H \m{X}_k^H \m{X}_k \hat{\ve{g}}_{k-1} \approx L \sigma_x^2 \operatorname{E}\left[||\hat{\ve{g}}_{k-1}||_2^2\right]$ and $\hat{\ve{h}}^H_{k-1} \m{X}_k \m{X}_k^H \hat{\ve{h}}_{k-1} \approx M \sigma_x^2 \operatorname{E}\left[||\hat{\ve{h}}_{k-1}||_2^2\right]$, and comparing \eqref{equ:111} and \eqref{equ:112} with \eqref{equ:15} and \eqref{equ:16}, it is easy to see that the boundaries for the normalized step-sizes result in
\begin{align}
	0 < \alpha_{\ve{h}} < 2
\end{align}
and
\begin{align}
	0 < \alpha_{\ve{g}} < 2 \text{.}
\end{align}
With the normalized step-sizes \eqref{equ:111} and \eqref{equ:112}, \eqref{equ:Delta} simplifies to $\Delta = 2 - (\alpha_{\ve{h}} + \alpha_{\ve{g}})$.
Clearly, $\Delta$ has to be greater than zero, which yields
\begin{align}
	\alpha_{\ve{h}} + \alpha_{\ve{g}} < 2
\end{align}
as an additional restriction for the normalized step-sizes.
Finally, \autoref{alg:nlmsalg} summarizes the \ac{cbnlms} filter. For even faster convergence, one can use a bilinear \ac{rls} filter, derived in the next section.

\begin{algorithm}[t]
	\DontPrintSemicolon
	\emph{Initialize variables:}\;
	$0 < \alpha_{\ve{h}} < 2$ and $0 < \alpha_{\ve{g}} < 2$ with $\alpha_{\ve{h}} + \alpha_{\ve{g}} < 2$\;
	$\delta_{\ve{h}} > 0$ and $\delta_{\ve{g}} > 0$\;
	$\hat{\ve{h}}_{0} \ne \mathbf{0}$ and $\hat{\ve{g}}_{0} \ne \mathbf{0}$\;
	\For{$k = 1,2,3,\ldots$}{
		$e_k = y_k - \hat{\ve{h}}_{k-1}^H \m{X}_k \hat{\ve{g}}_{k-1}$\;
		\emph{Parameter update of $\hat{\ve{h}}_{k}$:}\;
		$\hat{\ve{h}}_k = \hat{\ve{h}}_{k-1} + \frac{\alpha_{\ve{h}} \m{X}_k \hat{\ve{g}}_{k-1}}{\delta_{\ve{h}} + \hat{\ve{g}}_{k-1}^H \m{X}_k^H \m{X}_k \hat{\ve{g}}_{k-1}} e^*_k$\;
		\BlankLine		  \emph{Parameter update of $\hat{\ve{g}}_{k}$:}\;
		$\hat{\ve{g}}_k = \hat{\ve{g}}_{k-1} + \frac{\alpha_{\ve{g}} \m{X}_k^H \hat{\ve{h}}_{k-1}}{\delta_{\ve{g}} + \hat{\ve{h}}^H_{k-1} \m{X}_k \m{X}_k^H \hat{\ve{h}}_{k-1}} e_k$\;
	}
	\caption{\ac{cbnlms} filter}
	\label{alg:nlmsalg}
\end{algorithm}

\subsection{\ac{cbrls} filter} \label{sec:CVBRLS}
The cost function for the \ac{cbrls} is defined as
\begin{align}
	J_k = \sum_{i = 1}^{k} \lambda^{k-i} \bar{e}_i \bar{e}_i^* \label{equ:123456}
\end{align}
where $\bar{e}_i = y_i - \hat{\ve{h}}_k^H \m{X}_i \hat{\ve{g}}_{k}$ and $\lambda \in \mathbb{R}$ is the forgetting factor. As with the linear \ac{rls} filter, the forgetting factor should be chosen to balance tracking performance and steady-state error. In a slowly varying system environment it may be assumed that $\hat{\ve{h}}_k^H \m{X}_i \hat{\ve{g}}_{k} \approx \hat{\ve{h}}_k^H \m{X}_i \hat{\ve{g}}_{i-1}$ when the index $i$ is close to $k$. For $i \ll k$ the introduced error gets attenuated by $\lambda^{k-i}$ \cite{Gebhard_2019_1}. To derive the first update equation for $\hat{\ve{h}}_k$ of the \ac{cbrls} filter, the cost function is approximated as
\begin{align}
	J_k \approx \sum_{i = 1}^{k} \lambda^{k-i} \epsilon_i \epsilon_i^* \label{equ:007}
\end{align}
where $\epsilon_i = y_i - \hat{\ve{h}}_k^H \m{X}_i \hat{\ve{g}}_{i-1}$.
The gradient of \eqref{equ:007} with respect to $\hat{\ve{h}}_k^*$ yields
 \begin{align}
 	\frac{\partial{J_k}}{\partial{\hat{\ve{h}}_{k}^*}} &= \sum_{i = 1}^{k} \lambda^{k-i} \m{X}_i \hat{\ve{g}}_{i-1} \hat{\ve{g}}_{i-1}^H \m{X}_i^H \hat{\ve{h}}_{k} - \label{equ:gradrls} \\
 	& \quad \,\sum_{i = 1}^{k} \lambda^{k-i} y_i^* \m{X}_i \hat{\ve{g}}_{i-1} \nonumber   \text{.}
 \end{align}
 Introducing
 \begin{align}
 	\widetilde{\m{R}}_{\ve{g},k} &= \sum_{i = 1}^{k} \lambda^{k-i} \m{X}_i \hat{\ve{g}}_{i-1} \hat{\ve{g}}_{i-1}^H \m{X}_i^H \in \mathbb{C}^{L \times L} \label{equ:Rhk} \\
 	&= \m{X}_k \hat{\ve{g}}_{k-1} \hat{\ve{g}}_{k-1}^H \m{X}_k^H + \lambda \widetilde{\m{R}}_{\ve{g},k-1}
 \end{align}
 and
 \begin{align}
 	\widetilde{\ve{r}}_{\ve{g},k} &= \sum_{i = 1}^{k} \lambda^{k-i} y_i^* \m{X}_i \hat{\ve{g}}_{i-1} \in \mathbb{C}^{L} \label{equ:rhk} \\
 	&= y_k^* \m{X}_k \hat{\ve{g}}_{k-1} + \lambda \widetilde{\ve{r}}_{\ve{g},k-1} \text{,} \label{equ:17}
 \end{align}
 and setting \eqref{equ:gradrls} to zero yields
 \begin{align}
 	\hat{\ve{h}}_{k} = \widetilde{\m{R}}_{\ve{g},k}^{-1} \widetilde{\ve{r}}_{\ve{g},k} \text{.} \label{equ:uphrls}
 \end{align}
Utilizing Woodbury's matrix identity \cite{Woodbury_1950_1} and renaming $\widetilde{\m{R}}_{\ve{g},k}^{-1}$ as $\m{P}_{\ve{g},k}$ leads to
\begin{align}
	\m{P}_{\ve{g},k} &= \lambda^{-1} \m{P}_{\ve{g},k-1} -  \label{equ:008}\\
	& \quad \, \frac{\lambda^{-1} \m{P}_{\ve{g},k-1} \m{X}_k \hat{\ve{g}}_{k-1} \hat{\ve{g}}_{k-1}^H \m{X}_k^H \m{P}_{\ve{g},k-1}}{\lambda + \hat{\ve{g}}_{k-1}^H \m{X}_k^H \m{P}_{\ve{g},k-1} \m{X}_k \hat{\ve{g}}_{k-1}} \nonumber \text{.}
\end{align}
Subsequently, the gain vector
\begin{align}
	\ve{k}_{\ve{g},k} = \frac{\m{P}_{\ve{g},k-1} \m{X}_k \hat{\ve{g}}_{k-1}}{\lambda + \hat{\ve{g}}_{k-1}^H \m{X}_k^H \m{P}_{\ve{g},k-1} \m{X}_k \hat{\ve{g}}_{k-1}} \in \mathbb{C}^L \text{,} \label{equ:gainth}
\end{align}
can be defined. After a few reformulations, \eqref{equ:008} and \eqref{equ:gainth} can be simplified to
\begin{align}
	\m{P}_{\ve{g},k} &= \lambda^{-1} \left(\m{P}_{\ve{g},k-1} - \ve{k}_{\ve{g},k} \hat{\ve{g}}_{k-1}^H \m{X}_k^H \m{P}_{\ve{g},k-1}\right) \label{equ:upph}
\end{align}
and
\begin{align}
	\ve{k}_{\ve{g},k} &= \m{P}_{\ve{g},k} \m{X}_k \hat{\ve{g}}_{k-1} \label{equ:009} \text{,}
\end{align}
respectively.
Incorporating \eqref{equ:17} into \eqref{equ:uphrls} yields
\begin{align}
	\hat{\ve{h}}_{k} = y_k^* \m{P}_{\ve{g},k} \m{X}_k \hat{\ve{g}}_{k-1} + \lambda \m{P}_{\ve{g},k} \ve{r}_{\ve{g},k-1} \text{.}
\end{align}
With \eqref{equ:upph} and \eqref{equ:009} this can be rewritten to
\begin{align}
	\hat{\ve{h}}_{k} &= y_k^* \ve{k}_{\ve{g},k} +  \label{equ:45}\\
	&\quad \left(\m{P}_{\ve{g},k-1} - \ve{k}_{\ve{g},k} \hat{\ve{g}}_{k-1}^H \m{X}_k^H \m{P}_{\ve{g},k-1}\right) \ve{r}_{\ve{g},k-1} \text{.} \nonumber
\end{align}
Finally, \eqref{equ:45} can be simplified to
\begin{align}
	\hat{\ve{h}}_{k} = \hat{\ve{h}}_{k-1} + e_k^* \ve{k}_{\ve{g},k} \label{equ:fuph} \text{.}
\end{align}
The next step is to perform the same steps for $\hat{\ve{g}}_{k}$ by utilizing the assumption $\hat{\ve{h}}_k^H \m{X}_i \hat{\ve{g}}_{k} \approx \hat{\ve{h}}_{i-1}^H \m{X}_i \hat{\ve{g}}_{k}$. The approximated cost function yields
\begin{align}
	J_k \approx \sum_{i = 1}^{k} \lambda^{k-i} \epsilon_i \epsilon_i^* \label{equ:0007}
\end{align}
where $\epsilon_i = y_i - \hat{\ve{h}}_{i-1}^H \m{X}_i \hat{\ve{g}}_{k}$. The derivative of \eqref{equ:0007} with respect to $\hat{\ve{g}}_k^*$ yields
 \begin{align}
	\frac{\partial{J_k}}{\partial{\hat{\ve{g}}_{k}^*}} &= \sum_{i = 1}^{k} \lambda^{k-i} \m{X}_i^H \hat{\ve{h}}_{i-1} \hat{\ve{h}}_{i-1}^H \m{X}_i \hat{\ve{g}}_{k} - \\
	& \quad \, \sum_{i = 1}^{k} \lambda^{k-i} y_i \m{X}_i^H \hat{\ve{h}}_{i-1} \nonumber   \text{.}
\end{align}
Similarly to the derivation above, we obtain
\begin{align}
	\hat{\ve{g}}_{k} = \widetilde{\m{R}}_{\ve{h},k}^{-1} \widetilde{\ve{r}}_{\ve{h},k} \label{equ:011}
\end{align}
by introducing
\begin{align}
	\widetilde{\m{R}}_{\ve{h},k} &= \sum_{i = 1}^{k} \lambda^{k-i} \m{X}_i^H \hat{\ve{h}}_{i-1} \hat{\ve{h}}_{i-1}^H \m{X}_i \in \mathbb{R}^{M \times M} \\
	&= \m{X}_k^H \hat{\ve{h}}_{k-1} \hat{\ve{h}}_{k-1}^H \m{X}_k + \lambda \widetilde{\m{R}}_{\ve{h},k-1}
\end{align}
and
\begin{align}
	\widetilde{\ve{r}}_{\ve{h},k} &= \sum_{i = 1}^{k} \lambda^{k-i} y_i \m{X}_i^H \hat{\ve{h}}_{i-1} \in \mathbb{C}^{M} \\
	&= y_k \m{X}_k^H \hat{\ve{h}}_{k-1} + \lambda \widetilde{\ve{r}}_{\ve{h},k-1}^H \text{.} \label{equ:18}
\end{align}
Woodbury's matrix identity and renaming $\widetilde{\m{R}}_{\ve{h},k}^{-1}$ as $\m{P}_{\ve{h},k}$ produces
\begin{align}
	\m{P}_{\ve{h},k} &= \lambda^{-1} \m{P}_{\ve{h},k-1} -  \label{equ:pgk}\\
	& \quad \, \frac{\lambda^{-1} \m{P}_{\ve{h},k-1} \m{X}_k^H \hat{\ve{h}}_{k-1} \hat{\ve{h}}_{k-1}^H \m{X}_k \m{P}_{\ve{h},k-1}}{\lambda + \hat{\ve{h}}_{k-1}^H \m{X}_k \m{P}_{\ve{h},k-1} \m{X}_k^H \hat{\ve{h}}_{k-1}} \nonumber \text{.}
\end{align}
By introducing a second gain vector
\begin{align}
	\ve{k}_{\ve{h},k} = \frac{\m{P}_{\ve{h},k-1} \m{X}_k^H \hat{\ve{h}}_{k-1}}{\lambda + \hat{\ve{h}}_{k-1}^H \m{X}_k \m{P}_{\ve{h},k-1} \m{X}_k^H \hat{\ve{h}}_{k-1}} \in \mathbb{C}^M \text{,}\label{equ:gaingk}
\end{align}
\eqref{equ:pgk} can be rewritten as
\begin{align}
	\m{P}_{\ve{h},k} &= \lambda^{-1} \left(\m{P}_{\ve{h},k-1} - \ve{k}_{\ve{h},k} \hat{\ve{h}}_{k-1}^H \m{X}_k \m{P}_{\ve{h},k-1}\right) \label{equ:uppg} \text{.}
\end{align}
Furthermore, with \eqref{equ:uppg} the gain vector can be reformulated as
\begin{align}
	\ve{k}_{\ve{h},k} &= \m{P}_{\ve{h},k} \m{X}_k^H \hat{\ve{h}}_{k-1} \label{equ:010} \text{.}
\end{align}
Inserting \eqref{equ:uppg}, \eqref{equ:010} and \eqref{equ:18} into \eqref{equ:011} yields the second update equation
\begin{align}
	\hat{\ve{g}}_{k} &= \hat{\ve{g}}_{k-1} + e_k \ve{k}_{\ve{h},k} \label{equ:fupg} \text{,}
\end{align}
of the \ac{cbrls} filter. \\
\textit{Convergence:}

Assuming the approximations made in \eqref{equ:007} and \eqref{equ:0007} are valid, a rather intuitive convergence consideration can be made: Within one iteration step, the update of $\hat{\ve{h}}_k$ and $\hat{\ve{g}}_k$ cannot increase the costs, which yields a converging behavior. This consideration is supported by the simulation results presented in \autoref{sec:SIM}. However, due to these approximations, a mathematically rigorous proof of convergence might be not possible. Therefore, when applying the \ac{cbrls} in different applications, it is recommended referring to simulation results to evaluate convergence behavior.

In \autoref{alg:rlsalg}, the \ac{cbrls} filter is summarized.

\begin{algorithm}[t]
	\DontPrintSemicolon
	\emph{Initialize variables:}\;
	$\m{P}_{\ve{g},0} = \nu_\ve{g} \mathbf{I}^{L \times L}$ with $\nu_\ve{g}>0$\;
	$\m{P}_{\ve{h},0} = \nu_\ve{h} \mathbf{I}^{M \times M}$ with $\nu_\ve{h}>0$\;
	$0<\lambda\leq1$\;
	$\hat{\ve{h}}_{0} \ne \mathbf{0}$ and $\hat{\ve{g}}_{0} \ne \mathbf{0}$\;
	\For{$k = 1,2,3,\ldots$}{
		$e_k = y_k - \hat{\ve{h}}_{k-1}^H \m{X}_k \hat{\ve{g}}_{k-1}$\;
		\emph{Parameter update of $\hat{\ve{h}}_{k}$:}\;
		$\ve{k}_{\ve{g},k} = \frac{\m{P}_{\ve{g},k-1} \m{X}_k \hat{\ve{g}}_{k-1}}{\lambda + \hat{\ve{g}}_{k-1}^H \m{X}_k^H \m{P}_{\ve{g},k-1} \m{X}_k \hat{\ve{g}}_{k-1}}$\;
		$\m{P}_{\ve{g},k} = \lambda^{-1} \left(\m{P}_{\ve{g},k-1} - \ve{k}_{\ve{g},k} \hat{\ve{g}}_{k-1}^H \m{X}_k^H \m{P}_{\ve{g},k-1}\right)$\;
		$\hat{\ve{h}}_{k} = \hat{\ve{h}}_{k-1} + e_k^* \ve{k}_{\ve{g},k}$\;
		\BlankLine		  \emph{Parameter update of $\hat{\ve{g}}_{k}$:}\;
		$\ve{k}_{\ve{h},k} = \frac{\m{P}_{\ve{h},k-1} \m{X}_k^H \hat{\ve{h}}_{k-1}}{\lambda + \hat{\ve{h}}_{k-1}^H \m{X}_k \m{P}_{\ve{h},k-1} \m{X}_k^H \hat{\ve{h}}_{k-1}}$\;
		$\m{P}_{\ve{h},k} = \lambda^{-1} \left(\m{P}_{\ve{h},k-1} - \ve{k}_{\ve{h},k} \hat{\ve{h}}_{k-1}^H \m{X}_k \m{P}_{\ve{h},k-1}\right)$\;
		$\hat{\ve{g}}_{k} = \hat{\ve{g}}_{k-1} + e_k \ve{k}_{\ve{h},k}$\;
	}
	\caption{\ac{cbrls} filter}
	\label{alg:rlsalg}
\end{algorithm}

\subsection{Mixed complex-valued-real-valued bilinear filter structure} \label{sec:mixed}
For the sake of completeness, this section presents the results for the mixed complex-valued-real-valued bilinear filter structure. Instead of the system in \eqref{equ:bi}, a real-valued coefficient vector $\ve{g} \in \mathbb{R}^M$ is assumed. Except for the \ac{crbnlms} filter, the derivations of the following filters can be performed analogously to those above. Therefore, just the final update equations are presented in this section, while the derivation of the \ac{crbnlms} filter can be found in the appendix.\\

\textit{\Ac{crbwf}:}
\begin{align}
	\hat{\ve{h}}_n &= \m{R}_{\ve{g},n-1}^{-1} \m{R}_{\m{X}y} \hat{\ve{g}}_{n-1} \\
	\hat{\ve{g}}_n &= \operatorname{Re}\left[\m{R}_{\ve{h},n}\right]^{-1} \operatorname{Re}\left[\m{R}_{\m{X}y}^H \hat{\ve{h}}_n\right]
\end{align}\\

\textit{\Ac{crbls} filter:}
\begin{align}
	\hat{\ve{h}}_n &= \left(\sum_{i=1}^{N} \m{X}_i \hat{\ve{g}}_{n-1} \hat{\ve{g}}_{n-1}^H \m{X}_i^H\right)^{-1} \sum_{i=1}^{N} y_i^* \m{X}_i \hat{\ve{g}}_{n-1} \\
	\hat{\ve{g}}_n &= \left(\sum_{i=1}^{N} \operatorname{Re}\left[\m{X}_i^H \hat{\ve{h}}_{n} \hat{\ve{h}}_{n}^H \m{X}_i\right]\right)^{-1} \\
	&\qquad \sum_{i=1}^{N} \operatorname{Re}\left[y_i \m{X}_i^H \hat{\ve{h}}_{n} \right] \nonumber
\end{align}\\

\textit{\Ac{crblms} filter:}
\begin{align}
	\hat{\ve{h}}_k &= \hat{\ve{h}}_{k-1} + \mu_{\ve{h}} e_k^* \m{X}_k \hat{\ve{g}}_{k-1} \label{equ:20} \\
	\hat{\ve{g}}_k &=\hat{\ve{g}}_{k-1} + \mu_{\ve{g}} 2 \operatorname{Re}\left[e_k \m{X}_k^H \hat{\ve{h}}_{k-1}\right] \label{equ:23}
\end{align}\\

\textit{\Ac{crbnlms} filter:}
\begin{align}
	&\hat{\ve{h}}_k = \hat{\ve{h}}_{k-1} + \frac{\alpha_{\ve{h}} \m{X}_k \hat{\ve{g}}_{k-1}}{\delta_{\ve{h}} + \hat{\ve{g}}_{k-1}^T \m{X}_k^H \m{X}_k \hat{\ve{g}}_{k-1}} e_k^* \\
	&\hat{\ve{g}}_k = \hat{\ve{g}}_{k-1} + \alpha_{\ve{g}} 2 \operatorname{Re}\left[e_k \m{X}_k^H \hat{\ve{h}}_{k-1}\right]\\
	&\frac{\operatorname{Re}\left[e_k^* \hat{\ve{h}}_{k-1}^H \m{X}_k\right] \operatorname{Re}\left[e_k \m{X}_k^H \hat{\ve{h}}_{k-1}\right] }{\delta_{\ve{g}} + \operatorname{Re}\left[e_k \hat{\ve{h}}_{k-1}^T \m{X}_k^*\right] \m{X}_k^H \hat{\ve{h}}_{k-1} \hat{\ve{h}}_{k-1}^H \m{X}_k \operatorname{Re}\left[e_k \m{X}_k^H \hat{\ve{h}}_{k-1}\right]} \nonumber
\end{align}\\

\textit{\Ac{crbrls} filter:}
\begin{align}
	\ve{k}_{\ve{g},k} &= \frac{\m{P}_{\ve{g},k-1} \m{X}_k \hat{\ve{g}}_{k-1}}{\lambda + \hat{\ve{g}}_{k-1}^T \m{X}_k^H \m{P}_{\ve{g},k-1} \m{X}_k \hat{\ve{g}}_{k-1}} \\
	\m{K}_{\ve{h},k} &= \lambda^{-1} \tilde{\m{P}}_{\ve{h},k-1} \tilde{\m{X}}_k \left(\m{I}^{2 \times 2} + \vphantom{\tilde{\m{X}}_k^H \lambda^{-1} \tilde{\m{P}}_{\ve{h},k-1} \tilde{\m{X}}_k} \right. \\
	&\quad \left.\tilde{\m{X}}_k^H \lambda^{-1} \tilde{\m{P}}_{\ve{h},k-1} \tilde{\m{X}}_k\right)^{-1} \nonumber \\
\tilde{\m{P}}_{\ve{h},k} &= \m{P}_{\ve{h},k} + \m{P}_{\ve{h},k}^* \\
\tilde{\m{X}}_k &= \begin{bmatrix}
	\m{X}_k^H \hat{\ve{h}}_{k-1}	&	\m{X}_k^T \hat{\ve{h}}_{k-1}^*
\end{bmatrix} \\
	\m{P}_{\ve{g},k} &= \lambda^{-1} \left(\m{P}_{\ve{g},k-1} - \ve{k}_{\ve{g},k} \hat{\ve{g}}_{k-1}^T \m{X}_k^H \m{P}_{\ve{g},k-1}\right) \\
	\tilde{\m{P}}_{\ve{h},k} &= \lambda^{-1} \left(\tilde{\m{P}}_{\ve{h},k-1} - \m{K}_{\ve{h},k} \tilde{\m{X}}_k^H \tilde{\m{P}}_{\ve{h},k-1}\right) \\
	\hat{\ve{h}}_{k} &= \hat{\ve{h}}_{k-1} + e_k^* \ve{k}_{\ve{g},k} \\
	\hat{\ve{g}}_k &= \hat{\ve{g}}_{k-1} + \m{K}_{\ve{h},k} \underline{\ve{e}}_k
\end{align}

\section{SIMULATION RESULTS}\label{sec:SIM}
This section presents simulation results in the context of system identification. First, several simulations with complex-valued \ac{miso} systems similar as in \cite{Benesty_2017_1,Bai_2006_1,Ciochina_2017_1,Dogariu_2018_1,Paleologu_2017_1,EliseiIliescu_2017_1,EliseiIliescu_2018_2,EliseiIliescu_2018_1,EliseiIliescu_2018_3,Dogariu_2018_2} are performed. Second, the identifications of complex-valued Hammerstein systems are regarded.
\subsection{Identification of a complex-valued \ac{miso} system}\label{sec_MISO}
The input signal matrix is given by
 \begin{align}
	\m{X}_k = \begin{bmatrix}
		\ve{x}_k^T \\
		\ve{x}_{k-1}^T \\
		\vdots \\
		\ve{x}_{k-L+1}^T
	\end{bmatrix} \text{,}
\end{align}
where $\ve{x}_k = \begin{bmatrix}
	x_{1,k}	&	x_{2,k}	&	\cdots	&	x_{M,k}
\end{bmatrix}^T \in \mathbb{C}^M$ contains the $M$ input samples of the complex-valued \ac{miso} system, at a time instance $k$, as can be seen in \autoref{fig:MISO}.

\subsubsection{Bilinear versus linear approach}\label{sec_BVSL}
\begin{figure*}[!t]
\centering
\includegraphics[width=\linewidth]{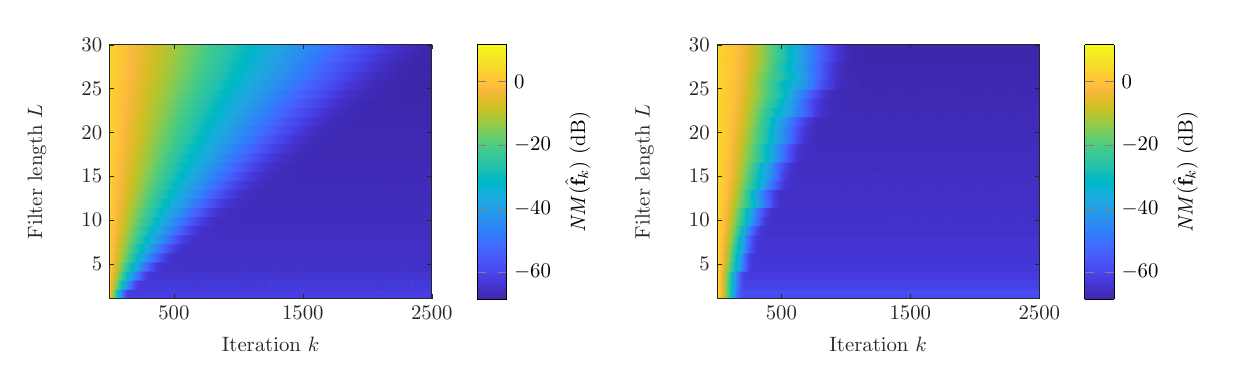}
\caption{The behavior of complex-valued NLMS algorithms when varying the filter length $L$. Left: \ac{clnlms}. Right: \ac{cbnlms}. The colors represent the normalized misalignment \eqref{equ:NM_meas} in dB.}
\label{fig:LMS_f12}
\end{figure*}

As a first task we regard the identification of a complex-valued \ac{miso} system of the form in \eqref{equ:bi}, using both the \ac{cbnlms} and the \ac{clnlms} filter.
For the following simulation, we chose $M=5$, proper complex-valued Gaussian input signals $x_{m,k} \sim \mathcal{CN}\left(0,1\right)$, and proper complex-valued Gaussian noise $n_k \sim \mathcal{CN}\left(0,1\right)$. The initial values $\hat{\ve{h}}_0$ and $\hat{\ve{g}}_0$, as well as the true values $\ve{h}$ and $\ve{g}$, are randomly chosen from a proper complex-valued normal distribution with zero mean and a standard deviation of $\sigma = 10$. To observe the influence of the filter length, $L$ has been increased from two to $30$. Note that the lengths of $\hat{\ve{h}}_k$ and $\hat{\ve{g}}_k$ were chosen to be equal to the lengths of $\ve{h}$ and $\ve{g}$, respectively. As explained in \eqref{equ:amb}, it is only possible to estimate $\ve{h}$ and $\ve{g}$ up to a complex scalar $\nu$. Therefore, the evaluation is performed on $\hat{\ve{f}}_k = \hat{\ve{g}}_k \otimes \hat{\ve{h}}_k^*$ using the normalized misalignment
\begin{align}
	\mathit{NM}\left(\hat{\ve{f}}_k\right) = \frac{||\ve{f} - \hat{\ve{f}}_k||_2^2}{||\ve{f}||_2^2} \in \mathbb{R} \text{.} \label{equ:NM_meas}
\end{align}
In \autoref{fig:LMS_f12}, the \ac{clnlms} filter (left) and the \ac{cbnlms} filter (right) are compared, where $\alpha_{\ve{h}} = \alpha_{\ve{g}} = 0.7$, $\alpha_{\ve{f}} = 1$, and $\delta_{\ve{h}} = \delta_{\ve{g}} = \delta_{\ve{f}} = 10^{-2}$. $\alpha_{\ve{f}}$ and $\delta_{\ve{f}}$ are the parameters of the \ac{clnlms} filter. All parameters were chosen to yield approximately the same steady-state performance for all involved filters and are summarized in \autoref{tab:A1}.
\begin{table*}[!t]
	\caption{Key simulation parameters for the fully complex-valued bilinear filters}
	\label{tab:A1}
	\centering
	\begin{tabular}{*{7}{c|}c}
		\hline
		simulation &	\multicolumn{2}{c|}{\ac{clnlms} filter}	&	\multicolumn{4}{c|}{\ac{cbnlms} filter}	&	\multicolumn{1}{c}{\ac{cbrls} filter}	\\
		&	$\alpha_{\ve{f}}$	&	$\delta_{\ve{f}}$	&	$\alpha_{\ve{h}}$	&	$\alpha_{\ve{g}}$	&	$\delta_{\ve{h}}$	&	$\delta_{\ve{g}}$	&	$\lambda$	\\
		\hline
		\autoref{sec_BVSL}	&	$1$	&	$10^{-2}$	&	$0.7$	&	$0.7$	&	$10^{-2}$	&	$10^{-2}$	& --	\\
		\autoref{sec_all}	&	--	&	--	&	$0.5$	&	$0.5$	&	$10^{-4}$	&	$10^{-4}$	&	$\frac{63}{64}$	\\
		\autoref{sec_all2}	&	--	&	--	&	$0.5$	&	$0.5$	&	$10^{-4}$	&	$10^{-4}$	&	$0.95$	\\		
		\hline
	\end{tabular}
\end{table*}
As $L$ increases, more iterations are required for convergence to the same level. However, it is evident that, especially for larger filter lengths, the bilinear filter converges much faster than the linear one. This is because the linear filter estimates $L M$ unknown complex-valued coefficients, whereas the bilinear filter just needs to estimate $L + M$ coefficients.

\subsubsection{\ac{cbwf} versus \ac{cbls} filter}
\begin{figure}[!t]
\centering
\includegraphics[width=\columnwidth]{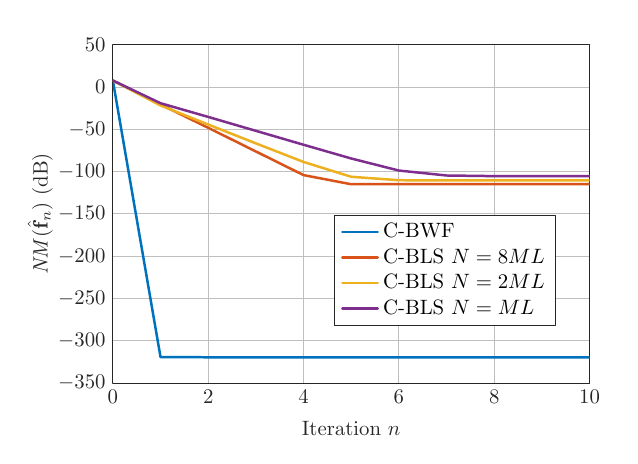}
\caption{Convergence curves of the \ac{cbwf} and the \ac{cbls} filter using proper complex-valued white Gaussian noise for the input signals.}
\label{fig:1_test}
\end{figure}

In this section, simulations involving the proposed \ac{cbwf} and the \ac{cbls} filter are presented. To demonstrate the behavior of those filters, we consider a filtering task similar to the one in the previous section. Note that for the \ac{cbwf}, it is necessary to know the statistical properties $\m{R}_{\widetilde{\ve{x}}\widetilde{\ve{x}}}$ and $\m{R}_{\m{X}y}$. If these properties are not available, they may be estimated in advance with \eqref{equ:12} and \eqref{equ:13}, or by using the \ac{cbls} filter. As above, a bilinear system of the form \eqref{equ:bi} with the same structure for the input signal matrix $\m{X}_k$, with $L = 64$ and $M = 5$, is considered. The unknown vectors $\ve{h}$ and $\ve{g}$, as well as the initial value $\hat{\ve{g}}_0$ were selected from the same random distribution as described above. For the first simulation in \autoref{fig:1_test}, the statistical properties were assumed to be $\m{R}_{\widetilde{\ve{x}}\widetilde{\ve{x}}} = \m{I}^{L M\times L M}$ and $\m{R}_{\m{X}y} = \operatorname{mat}_M\left[\m{R}_{\widetilde{\ve{x}}\widetilde{\ve{x}}} \ve{f}^*\right]$, where $x_{m,k}$ represents proper complex-valued white Gaussian noise. Since $\hat{\m{R}}_{\widetilde{\ve{x}}\widetilde{\ve{x}}}$ has to be invertible, the minimum number of data-points is $N = L M$. As described in \autoref{sec:FCVBF}, the \ac{cbwf}, with perfectly known statistics, converges in just one iteration step. However, one can see that the \ac{cbls} filter does not converge in one iteration step, and using more data improves the estimation of $\m{R}_{\widetilde{\ve{x}}\widetilde{\ve{x}}}$ and $\m{R}_{\m{X}y}$, resulting in faster and better convergence.

\begin{figure}[!t]
\centering
\includegraphics[width=\columnwidth]{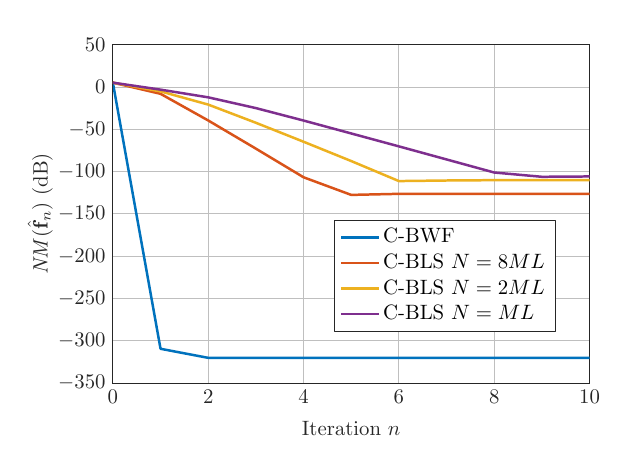}
\caption{Convergence curves of the \ac{cbwf} and the \ac{cbls} filter using a complex-valued moving average process for the input signals.}
\label{fig:2_test}
\end{figure}

In \autoref{fig:2_test}, instead of proper complex-valued white Gaussian noise, a complex-valued first-order moving average process is used. For this, proper complex-valued Gaussian random signals $u_{m,k} \sim \mathcal{CN}\left(0,0.5\right)$ are used to generate the input signals as $x_{m,k} = u_{m,k} + u_{m,k-1}$.
Similar to the previous case, the result improves as the estimation of the statistical properties becomes more accurate. Furthermore, as mentioned in \autoref{sec:FCVBF}, using the exact statistical properties leads to convergence in one iteration. Note that in \autoref{fig:2_test}, it might appear otherwise, which can be attributed to the simulation accuracy of 64-bit floating-point arithmetic.


\subsubsection{\ac{cbnlms} filter versus \ac{cbrls} filter versus \acs{2rbnlms} filter versus \acs{4rbnlms} filter}\label{sec_all}
\begin{figure}[!t]
\centering
\includegraphics[width=\columnwidth]{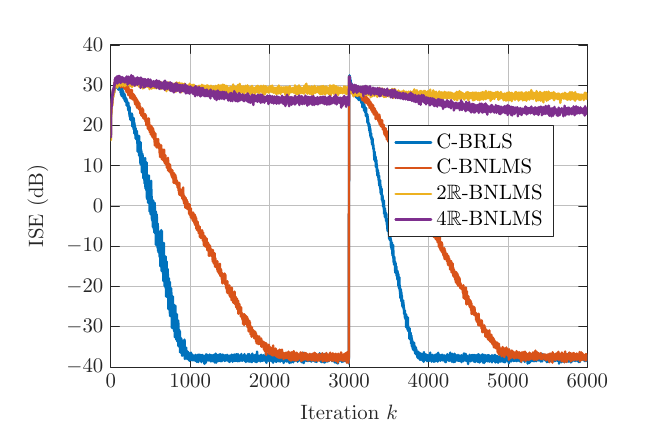}
\caption{Convergence behavior of the \ac{cbrls} filter, the \ac{cbnlms} filter, the \ac{2rbnlms} filter, and the \ac{4rbnlms} filter, when identifying a complex-valued \ac{miso} system.}
\label{fig:MISOComparison}
\end{figure}

In the next simulations we compare the \ac{cbrls} filter with the complex-valued bilinear \ac{nlms}-based filters. This is done by identifying the same complex-valued bilinear system as in the previous simulations. The input and noise signals are picked from proper complex-valued Gaussian distributions $x_{m,k} \sim \mathcal{CN}\left(0,\sigma_x^2\right)$, $n_k \sim \mathcal{CN}\left(0,\sigma_n^2\right)$ with $\sigma_n = \sigma_x = 0.01$. To implement the \ac{cbnlms} filter, the normalized step-sizes and the regularization constants are chosen as $\alpha_{\ve{h}} = \alpha_{\ve{g}} = 0.5$ and $\delta_{\ve{h}}=\delta_{\ve{g}} = 10^{-4}$, respectively. For the initialization of the \ac{cbrls} filter $\m{P}_{\ve{h},0} = 10 \m{I}^{M \times M}$, $\m{P}_{\ve{g},0} = 10 \m{I}^{L \times L}$, and $\lambda = 1 - \frac{1}{L}$ were selected. The latter values were picked similarly to those in \cite{EliseiIliescu_2017_1}, while the values for the \ac{cbnlms} filter were selected to ensure that both filters achieve about the same level of normalized misalignment. Furthermore, the parameters for the \ac{2rbnlms} and the \ac{4rbnlms} filters were selected as $\alpha_{\ve{h}_i} = \alpha_{\ve{g}_i} = 0.17$, $\delta_{\ve{h}_i} = \delta_{\ve{g}_i} = \delta_{\ve{h}_{\operatorname{Re}}} = \delta_{\ve{h}_{\operatorname{Im}}} = \delta_{\ve{g}_{\operatorname{Re}}} = \delta_{\ve{g}_{\operatorname{Im}}} = \delta_{\ve{h}}$, and $\alpha_{\ve{h}_{\operatorname{Re}}} = \alpha_{\ve{h}_{\operatorname{Im}}} = \alpha_{\ve{g}_{\operatorname{Re}}} = \alpha_{\ve{g}_{\operatorname{Im}}} = 0.15$, for $i = 1, \dots ,4$. As expected, the \ac{cbrls} filter is the fastest in convergence, as can be seen in \autoref{fig:MISOComparison}. Nevertheless, the \ac{cbnlms} filter also reaches around the same steady-state performance. Furthermore, since the unknown bilinear system cannot be appropriately modeled with the $2\mathbb{R}$ as well as the $4\mathbb{R}$ method, those filters failed to identify the system reasonably. Similar as in \cite{Dogariu_2018_1}, to evaluate the tracking abilities of the filters, an abrupt change in the parameter vectors $\ve{h}$ and $\ve{g}$ at $k = 3000$ was simulated. As expected, similar results were observed. The \ac{cbrls} filter adapts the fastest to the new parameter vectors, followed by the \ac{cbnlms} filter. The \ac{2rbnlms} filter and the \ac{4rbnlms} filter again fail to identify the system reasonably.
\begin{table*}[!t]
	\caption{Key simulation parameters for the \ac{2rbnlms} filter and the \ac{4rbnlms} filter}
	\label{tab:A2}
	\centering
	\begin{tabular}{*{12}{c|}c}
		\hline
		simulation &	\multicolumn{8}{c|}{\ac{2rbnlms} filter}	&	\multicolumn{4}{c}{\ac{4rbnlms} filter}	\\
		&	$\alpha_{\ve{h}_{\operatorname{Re}}}$	&	$\alpha_{\ve{h}_{\operatorname{Im}}}$	&	$\alpha_{\ve{g}_{\operatorname{Re}}}$	&	$\alpha_{\ve{g}_{\operatorname{Im}}}$	& $\delta_{\ve{h}_{\operatorname{Re}}}$	&	$\delta_{\ve{h}_{\operatorname{Im}}}$	&	$\delta_{\ve{g}_{\operatorname{Re}}}$	& $\delta_{\ve{g}_{\operatorname{Im}}}$	&	$\alpha_{\ve{h}_{i}}$	&	$\alpha_{\ve{g}_{i}}$	&	$\delta_{\ve{h}_{i}}$		&	$\delta_{\ve{g}_{i}}$	\\
		\hline
		
		\autoref{sec_all}	&	$0.15$	&	$0.15$	&	$0.15$	&	$0.15$	&	$10^{-4}$	&	$10^{-4}$	&	$10^{-4}$	&	$10^{-4}$	&	$0.17$	&	$0.17$	&	$10^{-4}$	&	$10^{-4}$	\\
		
		\autoref{sec_all2}	&	$1.7 \cdot 10^{-3}$	&	$1.7 \cdot 10^{-3}$	&	$1.7 \cdot 10^{-3}$	&	$1.7 \cdot 10^{-3}$	&	$10^{-4}$	&	$10^{-4}$	&	$10^{-4}$	&	$10^{-4}$	&	$10^{-4}$	&	$10^{-4}$	&	$10^{-4}$	&	$10^{-4}$	\\
		
		\hline
	\end{tabular}
\end{table*}

\subsection{Identification of a complex-valued Hammerstein system} \label{sec_Hammer}
The following simulation presents the identification of a complex-valued Hammerstein system, as illustrated in \autoref{fig:Hammerstein}.
\subsubsection{\ac{cbnlms} filter versus \ac{cbrls} filter versus \ac{2rbnlms} filter versus \ac{4rbnlms} filter} \label{sec_all2}
\begin{figure}[!t]
	\centering
	\includegraphics[width=\columnwidth]{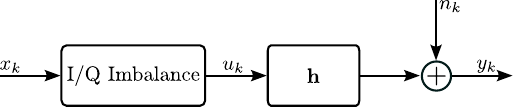}
	\caption{Block diagram of a communication transmitter in combination with a linear channel.}
	\label{fig:Hammerstein}
\end{figure}

\begin{figure}[!t]
\centering
\includegraphics[width=\columnwidth]{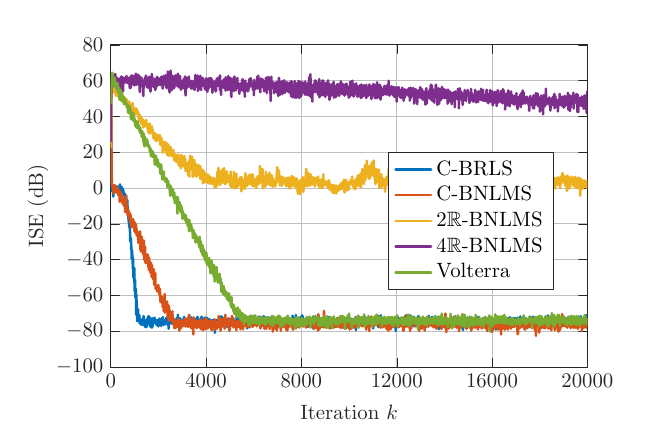}
\caption{Convergence behavior of the \acs{cbrls} filter, the \acs{cbnlms} filter, the \acs{2rbnlms} filter, the \acs{4rbnlms} filter, and the complex-valued \acs{nlms}-based Volterra filter, when identifying a complex-valued Hammerstein system.}
\label{fig:MISOComparison_Hammer1}
\end{figure}

In this section, we identify a communication transmitter impairment followed by a linear channel. In the complex-valued baseband, this cascade can be modeled as a complex-valued Hammerstein system: a memoryless front-end followed by a linear time-invariant channel, as shown in \autoref{fig:Hammerstein}.
In a typical transmitter, bits are mapped to complex-valued symbols, shaped by a pulse-shaping filter, upconverted by an \ac{iq} modulator, amplified by a \ac{pa}, and then transmitted over the channel. In the simplified example considered here, both the pulse shaper and the \ac{pa} are modeled as linear. The only memoryless distortion arises from \ac{iq} imbalance in the modulator, which maps the complex-valued baseband symbols to
\begin{align}
	u_k = g_1 x_k + g_2 x_k^*
\end{align}
with $g_1 = \frac{1 + g_{\text{T}} e^{-\text{j}\phi_{\text{T}}}}{2} \in \mathbb{C}$ and $g_2 = \frac{1 - g_{\text{T}} e^{\text{j}\phi_{\text{T}}}}{2} \in \mathbb{C}$ \cite{Valkama_2001_1, Tockner_2024_1}. The amplitude imbalance factor is denoted by $g_{\text{T}} \in \mathbb{R}$, whereas $\phi_{\text{T}} \in \mathbb{R}$ represents the phase imbalance. Now, this signal is convolved with the impulse response of the channel, which is modeled as a complex-valued \ac{fir} filter $\ve{h} \in \mathbb{C}^L$. The output signal is
\begin{align}
	y_k = \ve{h}^H \begin{bmatrix}
		x_k	&	x_k^* \\
		\vdots	&	\vdots	\\
		x_{k-L+1}	&	x_{k-L+1}^*
	\end{bmatrix} \begin{bmatrix}
	g_1 \\
	g_2
	\end{bmatrix} = \ve{h}^H \m{X}_k \ve{g} \label{equ:IQB} \text{.}
\end{align}
From \eqref{equ:IQB} it can be seen, that fully complex-valued bilinear filters are suitable structures for identifying such a system.
On the other hand, because in general \eqref{equ:IQB} cannot be represented with a $2\mathbb{R}$ bilinear filter or a $4\mathbb{R}$ bilinear filter, the identification with a $2\mathbb{R}$ bilinear filter or a $4\mathbb{R}$ bilinear filter is expected to perform worse. The channel impulse response $\ve{h}$ with the length $L = 64$ shall model a multipath propagation environment and has been chosen similar as in \cite{Hofbauer_2016_1}. The amplitude imbalance and the phase imbalance were set to $g_{\text{T}} = 1.15$ and $\phi_{\text{T}} = \frac{\pi}{18}$, respectively. For the \ac{cbnlms} filter, $\alpha_{\ve{g}} = \alpha_{\ve{h}} = 0.5$ and $\delta_{\ve{g}} = \delta_{\ve{h}} = 10^{-4}$, were chosen. Furthermore, for the \ac{4rbnlms} filter the parameters were set to $\alpha_{\ve{h}_i} = \alpha_{\ve{g}_i} = 0.0001$ and $\delta_{\ve{h}_i} = \delta_{\ve{g}_i} = \delta_{\ve{h}}$, for $i = 1, \dots ,4$. For the \ac{2rbnlms} filter $\alpha_{\ve{g}_{\operatorname{Re}}} = \alpha_{\ve{g}_{\operatorname{Im}}} = \alpha_{\ve{h}_{\operatorname{Re}}} = \alpha_{\ve{h}_{\operatorname{Im}}} = 0.0017$ and $\delta_{\ve{g}_{\operatorname{Re}}} = \delta_{\ve{g}_{\operatorname{Im}}} = \delta_{\ve{h}_{\operatorname{Re}}} = \delta_{\ve{h}_{\operatorname{Im}}} = \delta_{\ve{h}}$ were chosen. The parameters for the \ac{cbrls} filter were set to $\lambda = 0.95$, $\m{P}_{\ve{h},0} = 10 \m{I}^{M \times M}$, and $\m{P}_{\ve{g},0} = 10 \m{I}^{L \times L}$. Because the Hammerstein system in this example can also be identified with standard complex-valued adaptive filters, we use the complex Volterra series in \cite{CrespoCadenas_2017_1} to implement an \ac{nlms}-based Volterra filter, which serves as the complex-valued nonlinear baseline. The parameters for both, the \ac{cbrls} filter and the complex-valued Volterra filter were chosen such that they achieve approximately the same steady-state performance as the \ac{cbnlms} filter. The parameters for the \ac{2rbnlms} and the \ac{4rbnlms} were chosen to achieve a similar speed of convergence. For the input as well as for the noise signal, proper complex-valued Gaussian distributed noise with $x_k \sim \mathcal{CN}\left(0,1\right)$ and $n_k \sim \mathcal{CN}\left(0,0.0001^2\right)$, respectively, was chosen.
The \ac{ise} results of this simulation are shown in \autoref{fig:MISOComparison_Hammer1}. Using either a $2\mathbb{R}$ bilinear filter or $4\mathbb{R}$ bilinear filter yields a high \ac{ise}, while the other three filters achieve substantially lower \ac{ise} with comparable steady-state performance. As expected, the \ac{cbrls} filter is the fastest. Notably, the \ac{cbnlms} filter is faster than the complex-valued Volterra filter, likely due to its smaller number of parameters to estimate.

\section{CONCLUSION}\label{sec:CON}
In this work, we introduced several novel complex-valued bilinear filters for the application in modeling and identifying complex-valued bilinear systems. First, complex-valued $2\mathbb{R}$ and $4\mathbb{R}$ bilinear filters were discussed. Next, several fully complex-valued bilinear filters, including the \ac{cbwf}, the \ac{cbls} filter, the \ac{cblms} filter, the \ac{cbnlms} filter, and the \ac{cbrls} filter were introduced. In addition, the update equations for mixed complex-valued-real-valued bilinear filters were presented. Furthermore, for the \acp{wf} and the \ac{lms}-based filters, investigations on the convergence behavior were conducted. Those filters were used to identify complex-valued \ac{miso} systems and complex-valued Hammerstein models.

\appendix
\section{Proof of convergence for the \ac{cbwf}} \label{app:wf}
Assuming statistical independent and identically distributed columns $\ve{x}_{i,k}$ of $\m{X}_k$, the filter will converge after one iteration step, which is proven in the following. Utilizing this assumption the covariance matrix $\m{R}_{\widetilde{\ve{x}}\widetilde{\ve{x}}}$ can be written as a block-diagonal matrix
\begin{align}
	\m{R}_{\widetilde{\ve{x}}\widetilde{\ve{x}}} = \m{I}^{M \times M} \otimes \m{R}_{\ve{x} \ve{x}} \text{,} \label{equ:Rxx}
\end{align}
where $\m{R}_{\ve{x} \ve{x}} = \operatorname{E}\left[\ve{x}_{i,k}\ve{x}_{i,k}^H\right] \in \mathbb{C}^{L \times L}$.
Inserting \eqref{equ:Rxx} into \eqref{equ:10} yields
\begin{align}
	\m{R}_{\ve{g},n-1} = \sum_{m=1}^{M} |\hat{g}_{n-1,m}|^2 \m{R}_{\ve{x} \ve{x}} \label{equ:02} \text{,}
\end{align}
with $\hat{g}_{n-1,m}$ as the $m$th element of $\hat{\ve{g}}_{n-1}$.
Inserting \eqref{equ:equiv} into the definition of $\m{R}_{\ve{X}y}$ leads to
\begin{align}
	\m{R}_{\ve{X}y} &= \operatorname{mat}_M\left[\m{R}_{\widetilde{\ve{x}}\widetilde{\ve{x}}} \ve{f}^*\right] \\
	&= \begin{bmatrix}
		g_1^* \m{R}_{\ve{x} \ve{x}} \ve{h}	&	\cdots	&	g_M^* \m{R}_{\ve{x} \ve{x}} \ve{h}
	\end{bmatrix} \label{equ:03} \text{,}
\end{align}
with $g_m$ as the $m$th element of $\ve{g}$.
By inserting \eqref{equ:02} and \eqref{equ:03} into \eqref{equ:001}, one obtains
\begin{align}
	\hat{\ve{h}}_{n} = &\left(\sum_{i=1}^{M} |\hat{g}_{n-1,i}|^2 \m{R}_{\ve{x} \ve{x}}\right)^{-1} \label{equ:0002} \\
	&\left(\sum_{i=1}^{M} \hat{g}_{n-1,i} g_i^* \m{R}_{\ve{x} \ve{x}}\right) \ve{h} \text{.} \nonumber
\end{align}
This can be simplified to
\begin{align}
	\hat{\ve{h}}_n = \frac{\ve{g}^H \hat{\ve{g}}_{n-1}}{\hat{\ve{g}}_{n-1}^H \hat{\ve{g}}_{n-1}} \ve{h} \text{.} \label{equ:0005}
\end{align}
As next step we use $\hat{\ve{h}}_{n}$ to calculate
\begin{align}
	\hat{\ve{g}}_n = \m{R}_{\ve{h},n}^{-1} \m{R}_{\ve{X}y}^H \hat{\ve{h}}_{n} \text{.}
\end{align}
Inserting \eqref{equ:Rxx} into \eqref{equ:11} yields the diagonal matrix
\begin{align}
	\m{R}_{\ve{h}, n} = \hat{\ve{h}}_{n}^T \m{R}_{\ve{x} \ve{x}}^* \hat{\ve{h}}_{n}^* \m{I}^{M \times M} \text{.} \label{equ:0004}
\end{align}
With \eqref{equ:03} and \eqref{equ:0004} the estimate $\hat{\ve{g}}_n$ follows to
\begin{align}
	\hat{\ve{g}}_n = \frac{\ve{h}^H \m{R}_{\ve{x} \ve{x}} \hat{\ve{h}}_{n}}{\hat{\ve{h}}_{n}^H \m{R}_{\ve{x} \ve{x}} \hat{\ve{h}}_{n}} \ve{g} \text{.} \label{equ:0003}
\end{align}
Inserting \eqref{equ:0005} leads to
\begin{align}
	\hat{\ve{g}}_n = \frac{\hat{\ve{g}}_{n-1}^H \hat{\ve{g}}_{n-1} }{\hat{\ve{g}}_{n-1}^H \ve{g}} \ve{g}  \label{equ:0006} \text{.}
\end{align}
From \eqref{equ:0005} one can see that
\begin{align}
	\hat{\ve{h}}_{1} = \underbrace{\frac{\ve{g}^H \hat{\ve{g}}_{0}}{\hat{\ve{g}}_{0}^H \hat{\ve{g}}_{0}}}_{\nu} \ve{h}
\end{align}
is already a scaled version of the true but unknown vector $\ve{h}$. Moreover, since it is only possible to estimate $\ve{h}$ up to a complex-valued scalar $\nu$, this is already the best estimate possible. Furthermore, the first estimate
\begin{align}
	\hat{\ve{g}}_1 = \frac{1}{\nu^*} \ve{g}
\end{align}
is also already a scaled version of $\ve{g}$. Consequently, the Kronecker product produces the true corresponding linear coefficient vector:
\begin{align}
	\hat{\ve{g}}_1 \otimes \hat{\ve{h}}_1^* = \ve{f} \text{.}
\end{align}
The proof continues by showing that the sequences do not change for $n > 1$, i.e.,
$
\hat{\ve{h}}_{n} = \hat{\ve{h}}_{1}
$
and
$
\hat{\ve{g}}_n = \hat{\ve{g}}_{1}
$
for $n \ge 1$.
From \eqref{equ:0005}, it is easy to see that if $\hat{\ve{g}}_{n-1} = \frac{1}{\nu^*}\ve{g}$, then
\begin{align}
	\hat{\ve{h}}_{n} = \nu \ve{h} \text{.}
\end{align}
Similarly, from \eqref{equ:0006} one can see that
\begin{align}
	\hat{\ve{g}}_{n} = \frac{1}{\nu^*}\ve{g} \text{.}
\end{align}
Hence, convergence is reached after one iteration step.

Note that if $\m{R}_{\widetilde{\ve{x}}\widetilde{\ve{x}}}$ does not fulfill \eqref{equ:Rxx}, $\hat{\ve{h}}_1$ and $\hat{\ve{g}}_1$ would not just be scaled versions of $\ve{h}$ and $\ve{g}$. Because of this, in the general case, the \ac{cbwf} does not converge after one iteration step.
Similarly, if the statistical properties $\m{R}_{\widetilde{\ve{x}}\widetilde{\ve{x}}}$ or $\m{R}_{\ve{X}y}$ are not known and therefore have to be estimated, convergence in one iteration step is not possible.

\section{Proof of convergence for the \ac{cblms} filter} \label{app:lms}
In the following, similarly as in the real-valued case \cite{Ciochina_2017_1} and assuming \eqref{equ:Rxx} with $\m{R}_{\ve{x} \ve{x}} = \sigma_x^2 \m{I}^{L \times L}$, step-size boundaries will be derived in order to guarantee convergence. Therefore, with a scalar $\nu \in \mathbb{C}$, the error vectors
\begin{align}
	\Delta \ve{h}_{k} = \nu \ve{h} - \hat{\ve{h}}_k  \label{equ:2}
\end{align}
and
\begin{align}
	\Delta \ve{g}_{k} = \frac{1}{\nu^*} \ve{g} - \hat{\ve{g}}_k  \text{,}\label{equ:6}
\end{align}
are introduced. These errors show the difference between the current estimates $\hat{\ve{h}}_k$ and $\hat{\ve{g}}_k$ and scaled versions of the true vectors $\ve{h}$ and $\ve{g}$, respectively. Inserting \eqref{equ:nuph} into \eqref{equ:2} yields
\begin{align}
	\Delta \ve{h}_{k} = \Delta \ve{h}_{k-1} - \mu_{\ve{h}} e_k^* \m{X}_k \hat{\ve{g}}_{k-1} \text{.}
\end{align}
Calculating the second-order moment of $||\Delta \ve{h}_{k}||$ results in
\begin{align}
	\operatorname{E}\left[||\Delta \ve{h}_{k}||_2^2\right] &= \operatorname{E}\left[||\Delta \ve{h}_{k-1}||_2^2\right] - \label{equ:5} \\
	&\quad \; \, 2 \mu_{\ve{h}} \operatorname{E}\left[\operatorname{Re}\left[e_k \hat{\ve{g}}_{k-1}^H \m{X}_k^H \Delta \ve{h}_{k-1}\right]\right] + \nonumber \\
	&\quad \; \, \mu_{\ve{h}}^2 \operatorname{E}\left[|e_k|^2 \hat{\ve{g}}_{k-1}^H \m{X}_k^H \m{X}_k \hat{\ve{g}}_{k-1}\right] \nonumber \text{.}
\end{align}
For a more compact notation
\begin{align}
	a_{\ve{h},k} = \operatorname{E}\left[\operatorname{Re}\left[e_k \hat{\ve{g}}_{k-1}^H \m{X}_k^H \Delta \ve{h}_{k-1}\right]\right] \label{equ:7}
\end{align}
and
\begin{align}
	b_{\ve{h},k} = \operatorname{E}\left[|e_k|^2 \hat{\ve{g}}_{k-1}^H \m{X}_k^H \m{X}_k \hat{\ve{g}}_{k-1}\right] \label{equ:8}
\end{align}
are introduced, which simplifies \eqref{equ:5} to
\begin{align}
	\operatorname{E}\left[||\Delta \ve{h}_{k}||_2^2\right] &= \operatorname{E}\left[||\Delta \ve{h}_{k-1}||_2^2\right] - \label{equ:deltah}\\
	&\quad \; \,2\mu_{\ve{h}} a_{\ve{h},k} + \mu_{\ve{h}}^2 b_{\ve{h},k} \nonumber  \text{.}
\end{align}
By inserting \eqref{equ:ebi} and \eqref{equ:bi} into \eqref{equ:7}, and assuming statistical independence between the noise $n_k$ and the input signal matrix $\m{X}_k$, \eqref{equ:7} can be simplified to
\begin{align}
	a_{\ve{h},k} = &\operatorname{Re}\left[\operatorname{E}\left[\ve{h}^H \m{X}_k \ve{g} \hat{\ve{g}}_{k-1}^H \m{X}_k^H \Delta \ve{h}_{k-1}\right] - \vphantom{\operatorname{E}\left[\hat{\ve{h}}_{k-1}^H \m{X}_k \hat{\ve{g}}_{k-1} \hat{\ve{g}}_{k-1} \m{X}_k^H \Delta \ve{h}_{k-1}\right]} \right. \\
	&\left.\operatorname{E}\left[\hat{\ve{h}}_{k-1}^H \m{X}_k \hat{\ve{g}}_{k-1} \hat{\ve{g}}_{k-1} \m{X}_k^H \Delta \ve{h}_{k-1}\right]\right] \nonumber \text{.}
\end{align}
Furthermore, using reformulated versions of \eqref{equ:2} and \eqref{equ:6} leads to
\begin{align}
	a_{\ve{h},k} = &\operatorname{Re}\left[\operatorname{E}\left[ \nu^* \ve{h}^H \m{X}_k \Delta \ve{g}_{k-1} \hat{\ve{g}}_{k-1}^H \m{X}_k^H \Delta \ve{h}_{k-1}\right] + \vphantom{\operatorname{E}\left[\Delta \ve{h}_{k-1}^H \m{X}_k \hat{\ve{g}}_{k-1} \hat{\ve{g}}_{k-1} \m{X}_k^H \Delta \ve{h}_{k-1}\right]} \right. \label{equ:123}\\
	&\left.\operatorname{E}\left[\Delta \ve{h}_{k-1}^H \m{X}_k \hat{\ve{g}}_{k-1} \hat{\ve{g}}_{k-1} \m{X}_k^H \Delta \ve{h}_{k-1}\right]\right] \text{.} \nonumber
\end{align}
Utilizing the property $\ve{v}^H \m{A} \ve{v} = \operatorname{tr}\left[\ve{v} \ve{v}^H \m{A}\right]$, the second part of the right-hand side of \eqref{equ:123} can be rewritten as
\begin{align}
	a_{\ve{h},k} = &\operatorname{Re}\left[\operatorname{E}\left[ \nu^* \ve{h}^H \m{X}_k \Delta \ve{g}_{k-1} \hat{\ve{g}}_{k-1}^H \m{X}_k^H \Delta \ve{h}_{k-1}\right] + \vphantom{\operatorname{E}\left[\Delta \ve{h}_{k-1}^H \m{X}_k \hat{\ve{g}}_{k-1} \hat{\ve{g}}_{k-1} \m{X}_k^H \Delta \ve{h}_{k-1}\right]} \right. \\
	&\left. \operatorname{tr}\left[\operatorname{E}\left[\Delta \ve{h}_{k-1} \Delta \ve{h}_{k-1}^H \m{X}_k \hat{\ve{g}}_{k-1} \hat{\ve{g}}_{k-1} \m{X}_k^H\right]\right]\right] \text{.} \nonumber
\end{align}
Furthermore, assuming $\Delta \ve{h}_{k-1}$ is uncorrelated to $\m{X}_k \hat{\ve{g}}_{k-1}$, yields
\begin{align}
	a_{\ve{h},k} = &\operatorname{Re}\left[\operatorname{E}\left[ \nu^* \ve{h}^H \m{X}_k \Delta \ve{g}_{k-1} \hat{\ve{g}}_{k-1}^H \m{X}_k^H \Delta \ve{h}_{k-1}\right] + \vphantom{\operatorname{E}\left[\Delta \ve{h}_{k-1}^H \m{X}_k \hat{\ve{g}}_{k-1} \hat{\ve{g}}_{k-1} \m{X}_k^H \Delta \ve{h}_{k-1}\right]} \right. \\
	&\left. \operatorname{tr}\left[\operatorname{E}\left[\Delta \ve{h}_{k-1} \Delta \ve{h}_{k-1}^H \right] \m{R}_{\ve{g},k-1}\right]\right] \text{.} \nonumber
\end{align}
Now, by incorporating \eqref{equ:Rxx} with $\m{R}_{\ve{x} \ve{x}} = \sigma_x^2 \m{I}^{L \times L}$, this can be further simplified to
\begin{align}
	a_{\ve{h},k} &= \operatorname{Re}\left[\nu^* \sigma_x^2 \ve{h}^H \operatorname{E}\left[\Delta \ve{h}_{k-1}\right] \operatorname{E}\left[\hat{\ve{g}}_{k-1}^H \Delta \ve{g}_{k-1} \right] \right] +  \label{equ:ah}\\
	&\quad \; \,\sigma_x^2 \operatorname{E}\left[||\hat{\ve{g}}_{k-1}||_2^2\right] \operatorname{E}\left[||\Delta \ve{h}_{k-1}||_2^2\right] \nonumber \text{.}
\end{align}
Using the approximation $\hat{\ve{g}}_{k-1}^H \m{X}_k^H \m{X}_k \hat{\ve{g}}_{k-1} \approx L \sigma_x^2 \operatorname{E}\left[||\hat{\ve{g}}_{k-1}||_2^2\right]$, \eqref{equ:8} can be rewritten to
\begin{align}
	b_{\ve{h},k} = L \sigma_x^2 \operatorname{E}\left[||\hat{\ve{g}}_{k-1}||_2^2\right] \operatorname{E}\left[|e_k|^2\right] \text{.} \label{equ:01234}
\end{align}
To simplify this, it can be shown that the error from \eqref{equ:ebi} can be written as
\begin{align}
	e_k = n_k + \lambda^* \ve{h}^H \m{X}_k \Delta \ve{g}_{k-1} + \Delta \ve{h}_{k-1} \m{X}_k \hat{\ve{g}}_{k-1} \text{.} \label{equ:0123}
	\end{align}
Inserting \eqref{equ:0123} into \eqref{equ:01234} together with similar reformulations as above yields
\begin{align}
	b_{\ve{h},k} &= L \sigma_x^2 \operatorname{E}\left[||\hat{\ve{g}}_{k-1}||_2^2\right] \left(\sigma_n^2 + \vphantom{|\nu|^2 \sigma_x^2 ||\ve{h}||_2^2 \operatorname{E}\left[||\Delta \ve{g}_{k-1}||_2^2\right] + 2\operatorname{Re}\left[\nu \sigma_x^2 \operatorname{E}\left[\Delta \ve{h}_{k-1}^H\right] \ve{h} \operatorname{E}\left[\Delta \ve{g}_{k-1}^H \hat{\ve{g}}_{k-1}\right]\right]} \right.  \label{equ:bh}\\
	&\quad \, \,|\nu|^2 \sigma_x^2 ||\ve{h}||_2^2 \operatorname{E}\left[||\Delta \ve{g}_{k-1}||_2^2\right] + \nonumber \\
	&\quad \,\left.2\operatorname{Re}\left[\nu \sigma_x^2 \operatorname{E}\left[\Delta \ve{h}_{k-1}^H\right] \ve{h} \operatorname{E}\left[\Delta \ve{g}_{k-1}^H \hat{\ve{g}}_{k-1}\right]\right] + \right. \nonumber \\
	&\quad \,\left.\sigma_x^2 \operatorname{E}\left[||\hat{\ve{g}}_{k-1}||_2^2\right] \operatorname{E}\left[||\Delta \ve{h}_{k-1}||_2^2\right] \right) \text{.}\nonumber
\end{align}
Inserting \eqref{equ:ah} and \eqref{equ:bh} into \eqref{equ:deltah} yields
\begin{align}
	\operatorname{E}\left[||\Delta \ve{h}_{k}||_2^2\right] = &\operatorname{E}\left[||\Delta \ve{h}_{k-1}||_2^2\right] c_{\ve{h},k} + d_{\ve{h},k}
\end{align}
where
\begin{align}
	c_{\ve{h},k} &= 1 - 2 \mu_{\ve{h}} \sigma_x^2 \operatorname{E}\left[||\hat{\ve{g}}_{k-1}||_2^2\right] + \\
	& \quad \, \mu_{\ve{h}}^2 \sigma_x^4 L \operatorname{E}\left[||\hat{\ve{g}}_{k-1}||_2^2\right]^2 \nonumber
\end{align}
and
\begin{align}
	d_{\ve{h},k} &= \mu_{\ve{h}}^2 L \sigma_x^2 \operatorname{E}\left[||\hat{\ve{g}}_{k-1}||_2^2\right] \left(\sigma_n^2 + \vphantom{2 \operatorname{Re}\left[\nu \sigma_x^2 \operatorname{E}\left[\Delta \ve{h}_{k-1}\right]^H \ve{h} \operatorname{E}\left[\Delta \ve{g}_{k-1}^H \hat{\ve{g}}_{k-1}\right]\right]} \right. \\
	&\quad\left. |\nu|^2 \sigma_x^2 ||\ve{h}||_2^2 \operatorname{E}\left[||\Delta \ve{g}_{k-1}||_2^2\right] + \right. \nonumber \\
	&\quad\,\left. 2 \operatorname{Re}\left[\nu \sigma_x^2 \operatorname{E}\left[\Delta \ve{h}_{k-1}\right]^H \ve{h} \operatorname{E}\left[\Delta \ve{g}_{k-1}^H \hat{\ve{g}}_{k-1}\right]\right] \right) - \nonumber \\
	&\quad \; \,2 \mu_{\ve{h}} \operatorname{Re}\left[\nu^* \sigma_x^2 \ve{h}^H \operatorname{E}\left[\Delta \ve{h}_{k-1}\right] \operatorname{E}\left[\hat{\ve{g}}_{k-1} \Delta \ve{g}_{k-1}\right] \right] \nonumber \text{.}
\end{align}
Note that $c_{\ve{h},k} > 0$, and to guarantee stability of the algorithm $c_{\ve{h},k} < 1$ should hold. $c_{\ve{h},k} < 1$ yields
\begin{align}
	- 2 \mu_{\ve{h}} \sigma_x^2 \operatorname{E}\left[||\hat{\ve{g}}_{k-1}||_2^2\right] + \mu_{\ve{h}}^2 \sigma_x^4 L \operatorname{E}\left[||\hat{\ve{g}}_{k-1}||_2^2\right]^2 < 0 \text{.}
\end{align}
Solving this quadratic inequality for $\mu_{\ve{h}}$, the boundaries
\begin{align}
	0 < \mu_{\ve{h}} < \frac{2}{L \sigma_x^2 \operatorname{E}\left[||\hat{\ve{g}}_{k-1}||_2^2\right]} \label{equ:15}
\end{align}
can be found.
With similar arguments,
\begin{align}
	\operatorname{E}\left[||\Delta \ve{g}_{k}||_2^2\right] = &\operatorname{E}\left[||\Delta \ve{g}_{k-1}||_2^2\right] c_{\ve{g},k} + d_{\ve{g},k}
\end{align}
can be derived, where
\begin{align}
	c_{\ve{g},k} &= 1 - 2 \mu_{\ve{g}} \sigma_x^2 \operatorname{E}\left[||\hat{\ve{h}}_{k-1}||_2^2\right] + \\
	&\quad \; \mu_{\ve{g}}^2 \sigma_x^4 M \operatorname{E}\left[||\hat{\ve{h}}_{k-1}||_2^2\right]^2 \nonumber
\end{align}
and
\begin{align}
	d_{\ve{g},k} &= \mu_{\ve{g}}^2 M \sigma_x^2 \operatorname{E}\left[||\hat{\ve{h}}_{k-1}||_2^2\right] \left(\sigma_n^2 + \vphantom{2 \operatorname{Re}\left[\frac{1}{\nu} \sigma_x^2 \operatorname{E}\left[\hat{\ve{h}}_{k-1}^H \Delta \ve{h}_{k-1}\right] \ve{g}^H \operatorname{E}\left[\Delta \ve{g}_{k-1}\right]\right]}\right. \\
	&\quad\left. \frac{1}{|\nu|^2} \sigma_x^2 ||\ve{g}||_2^2 \operatorname{E}\left[||\Delta \ve{h}_{k-1}||_2^2\right] + \right. \nonumber \\
	&\quad\,\left. 2 \operatorname{Re}\left[\frac{1}{\nu} \sigma_x^2 \operatorname{E}\left[\hat{\ve{h}}_{k-1}^H \Delta \ve{h}_{k-1}\right] \ve{g}^H \operatorname{E}\left[\Delta \ve{g}_{k-1}\right]\right] \right) - \nonumber \\
	&\quad \; \,2 \mu_{\ve{g}} \operatorname{Re}\left[\nu^* \sigma_x^2 \ve{h}^H \operatorname{E}\left[\Delta \ve{h}_{k-1}\right] \operatorname{E}\left[\hat{\ve{g}}_{k-1} \Delta \ve{g}_{k-1}\right] \right] \nonumber \text{.}
\end{align}
Again, it can be seen that $c_{\ve{g},k} > 0$, and for stability, $c_{\ve{g},k} < 1$ should hold. Similar to above, $c_{\ve{g},k} < 1$ yields
\begin{align}
	 - 2 \mu_{\ve{g}} \sigma_x^2 \operatorname{E}[||\hat{\ve{h}}_{k-1}||_2^2] + \mu_{\ve{g}}^2 \sigma_x^4 M \operatorname{E}[||\hat{\ve{h}}_{k-1}||_2^2]^2 < 0 \text{,}
\end{align}
which is again a quadratic inequality in $\mu_{\ve{g}}$. Solving this, results in the boundaries
\begin{align}
	0 < \mu_{\ve{g}} < \frac{2}{M \sigma_x^2 \operatorname{E}\left[||\hat{\ve{h}}_{k-1}||_2^2\right]} \text{.} \label{equ:16}
\end{align}
To analyze the algorithm after convergence, the terms
\begin{align}
	\operatorname{E}\left[||\Delta \ve{h}_{\infty}||_2^2\right] = \lim\limits_{k \rightarrow \infty} \operatorname{E}\left[||\Delta \ve{h}_{k}||_2^2\right]
\end{align}
and
\begin{align}
	\operatorname{E}\left[||\Delta \ve{g}_{\infty}||_2^2\right] = \lim\limits_{k \rightarrow \infty} \operatorname{E}\left[||\Delta \ve{g}_{k}||_2^2\right]
\end{align}
will be investigated.
Assuming appropriate step-sizes,
\begin{align}
	\operatorname{E}\left[||\Delta \ve{h}_{\infty}||_2^2\right] &= \lim\limits_{k \rightarrow \infty} \frac{d_{\ve{h},k}}{1 - c_{\ve{h},k}} \label{equ:14}\\
	&= \frac{\mu_{\ve{h}} L \sigma_n^2 + \mu_{\ve{h}} L \sigma_x^2 |\nu|^2 ||\ve{h}||_2^2 \operatorname{E}\left[||\Delta \ve{g}_{\infty}||_2^2\right]}{2 - \mu_{\ve{h}} \sigma_x^2 L \frac{1}{|\nu|^2}||\ve{g}||_2^2} \nonumber
\end{align}
and
\begin{align}
	\operatorname{E}\left[||\Delta \ve{g}_{\infty}||_2^2\right] &= \lim\limits_{k \rightarrow \infty} \frac{d_{\ve{g},k}}{1 - c_{\ve{g},k}} \label{equ:9} \\
	&= \frac{\mu_{\ve{g}} M \sigma_n^2 + \mu_{\ve{g}} M \sigma_x^2 \frac{1}{|\nu|^2} ||\ve{g}||_2^2 \operatorname{E}\left[||\Delta \ve{h}_{\infty}||_2^2\right]}{2 - \mu_{\ve{g}} \sigma_x^2 M |\nu|^2 ||\ve{h}||_2^2} \nonumber
\end{align}
can be obtained.
With \eqref{equ:14}, \eqref{equ:9}, and some additional reformulations
\begin{align}
	\operatorname{E}\left[||\Delta \ve{h}_{\infty}||_2^2\right] = \frac{\mu_{\ve{h}} L \sigma_n^2}{\Delta}
\end{align}
and
\begin{align}
	\operatorname{E}\left[||\Delta \ve{g}_{\infty}||_2^2\right] = \frac{\mu_{\ve{g}} M \sigma_n^2}{\Delta} \text{,}
\end{align}
with $\Delta = 2 - \sigma_x^2\left(\mu_{\ve{h}} L \frac{1}{|\nu|^2} ||\ve{g}||_2^2 + \mu_{\ve{g}} M |\nu|^2 ||\ve{h}||_2^2\right)$, can be derived.
Clearly, $\Delta$ has to be greater than zero, which yields an additional restriction for the step-sizes in order to guarantee convergence on the mean.

\section{Derivation of the \ac{crbnlms} filter}\label{sec:CVRVBLNLMS}
In this section, one possible method of how to normalize the step-size of the \ac{crblms} filter is shown. Similar as in \autoref{sec:CVBNLMS} for the \ac{cbnlms} filter, the a posteriori errors
\begin{align}
	\bar{e}_k &= y_k - \hat{\ve{h}}^H_k \m{X}_k \hat{\ve{g}}_{k-1} \label{equ:21}
\end{align}
and
\begin{align}
	\tilde{e}_k &= y_k - \hat{\ve{h}}^H_{k-1} \m{X}_k \hat{\ve{g}}_{k} \label{equ:22}
\end{align}
are introduced. Inserting \eqref{equ:20} into \eqref{equ:21} yields
\begin{align}
	\bar{e}_k &= e_k \left(1 - \mu_{\ve{h}} \hat{\ve{g}}_{k-1}^H \m{X}_k^H \m{X}_k \hat{\ve{g}}_{k-1}\right) \text{.}
\end{align}
Setting this error to zero and assuming $e_k \ne 0$, we obtain
\begin{align}
	\mu_{\ve{h}} = \frac{1}{\hat{\ve{g}}_{k-1}^H \m{X}_k^H \m{X}_k \hat{\ve{g}}_{k-1}} \text{.}
\end{align}
Using similar arguments as in \autoref{sec:CVBNLMS}, the final update equation for $\hat{\ve{h}}_k$ follows as
\begin{align}
	\hat{\ve{h}}_k &= \hat{\ve{h}}_{k-1} + \frac{\alpha_{\ve{h}} \m{X}_k \hat{\ve{g}}_{k-1}}{\delta_{\ve{h}} + \hat{\ve{g}}_{k-1}^T \m{X}_k^H \m{X}_k \hat{\ve{g}}_{k-1}} e_k^* \text{.}
\end{align}
Since the a posteriori error in \eqref{equ:22} is generally a complex-valued number, it might not be possible to set it to zero using a real-valued update for $\hat{\ve{g}}_k$. Setting \eqref{equ:22} to zero would require a complex-valued step-size $\mu_{\ve{g}}$. A meaningful step-size is the one that minimizes $\left|\tilde{e}_k\right|^2$. With \eqref{equ:22}, the following holds:
\begin{align}
	\left|\tilde{e}_k\right|^2 &= y_k y_k^* - 2 \operatorname{Re}\left[y_k^* \hat{\ve{h}}_{k-1}^H \m{X}_k\right] \hat{\ve{g}}_k + \\
	&\quad \, \, \hat{\ve{g}}_k^T \m{X}_k \hat{\ve{h}}_{k-1} \hat{\ve{h}}_{k-1}^H \m{X}_k \hat{\ve{g}}_k  \nonumber \text{.}
\end{align}
After inserting \eqref{equ:23} the gradient follows to
\begin{align}
	&\frac{\partial \left|\tilde{e}_k\right|^2}{\partial \mu_{\ve{g}}} = - 2 \operatorname{Re}\left[e_k^* \hat{\ve{h}}_{k-1}^H \m{X}_k\right] \operatorname{Re}\left[e_k \m{X}_k^H \hat{\ve{h}}_{k-1}\right] + \\
	& 2 \mu_{\ve{g}} \operatorname{Re}\left[e_k \hat{\ve{h}}_{k-1}^T \m{X}_k^*\right] \m{X}_k^H \hat{\ve{h}}_{k-1} \hat{\ve{h}}_{k-1}^H \m{X}_k \operatorname{Re}\left[e_k \m{X}_k^H \hat{\ve{h}}_{k-1}\right] \text{.} \nonumber
\end{align}
Setting this to zero yields the optimal step-size
\begin{align}
	&\mu_{\ve{g}} = \operatorname{Re}\left[e_k^* \hat{\ve{h}}_{k-1}^H \m{X}_k\right] \\
	&\frac{\operatorname{Re}\left[e_k \m{X}_k^H \hat{\ve{h}}_{k-1}\right]}{\operatorname{Re}\left[e_k \hat{\ve{h}}_{k-1}^T \m{X}_k^*\right] \m{X}_k^H \hat{\ve{h}}_{k-1} \hat{\ve{h}}_{k-1}^H \m{X}_k \operatorname{Re}\left[e_k \m{X}_k^H \hat{\ve{h}}_{k-1}\right]} \text{.} \nonumber
\end{align}
Again similar as in \autoref{sec:CVBNLMS}, the second update equation results in
\begin{align}
	&\hat{\ve{g}}_k = \hat{\ve{g}}_{k-1} + \alpha_{\ve{g}} 2 \operatorname{Re}\left[e_k \m{X}_k^H \hat{\ve{h}}_{k-1}\right]\\
	&\frac{\operatorname{Re}\left[e_k^* \hat{\ve{h}}_{k-1}^H \m{X}_k\right] \operatorname{Re}\left[e_k \m{X}_k^H \hat{\ve{h}}_{k-1}\right] }{\delta_{\ve{g}} + \operatorname{Re}\left[e_k \hat{\ve{h}}_{k-1}^T \m{X}_k^*\right] \m{X}_k^H \hat{\ve{h}}_{k-1} \hat{\ve{h}}_{k-1}^H \m{X}_k \operatorname{Re}\left[e_k \m{X}_k^H \hat{\ve{h}}_{k-1}\right]} \nonumber \text{.}
\end{align}

\bibliographystyle{elsarticle-num-names}
\bibliography{bibliography}

\end{document}